\documentclass[prb,twocolumn,superscriptaddress]{revtex4}

\usepackage[dvipdfmx]{graphicx}
\usepackage{dcolumn}
\usepackage{bm}
\usepackage{color}

\graphicspath{{fig/}}

\begin{document}

\preprint{APS/123-QED}

\title{
Designing nickelate superconductors with $d^8$ configuration exploiting 
mixed-anion strategy
} 
\author{Naoya Kitamine}
\author{Masayuki Ochi}
\author{Kazuhiko Kuroki}
\affiliation{Department of Physics, Osaka University, 1-1 Machikaneyama-cho, Toyonaka, Osaka, 560-0043, Japan}

\date{\today}

\begin{abstract}
Inspired by a recently proposed superconducting mechanism for a new cuprate superconductor Ba$_2$CuO$_{3+\delta}$, we theoretically design an unconventional nickelate superconductor with a $d^8$ electron configuration. Our strategy is to enlarge the on-site energy difference between $3d_{x^2-y^2}$ and other $3d$ orbitals by adopting halogens or hydrogen as out-of-plane anions, so that the $3d$ bands other than $d_{x^-y^2}$ lie just below the Fermi level for the $d^8$ configuration, acting as incipient bands that enhance superconductivity. We also discuss a possible relevance of the present proposal to the recently discovered superconductor (Nd,Sr)NiO$_2$.
\end{abstract}

\pacs{ }
\maketitle

The two families of high-$T_c$ superconductors, cuprates\cite{CuReview} and iron-based\cite{FeReview}, are often contrasted as single-orbital vs. multiorbital systems. Namely, in the cuprates, only the $d_{x^2-y^2}$ orbital plays the main role while in the iron-based superconductors, $d_{xy}$, $d_{xz}$, and $d_{yz}$ orbitals largely contribute to the electronic structure around the Fermi level ($E_F$). Recently, however, there has occurred a possible paradigm shift to this picture, owing to some new experimental findings. One such experiment is the discovery of high-$T_c$ superconductivity in Ba$_2$CuO$_{3+\delta}$\cite{Li}, where the apical oxygen height is so low that the energy levels of $d_{x^2-y^2}$ and $d_{3z^2-r^2}$ can be reversed compared to conventional cuprates.
Another example is a highly overdoped CuO$_2$ monolayer grown on $\mathrm{Bi}_{2}\mathrm{Sr}_{2}\mathrm{CaCu}_{2}\mathrm{O}_{8+\delta}$\cite{XueCuO2}, where $E_F$ is significantly lowered so that it reaches the $d_{3z^2-r^2}$ band \cite{WangCuO2}. 

\begin{figure}
	\includegraphics[width=8.5cm]{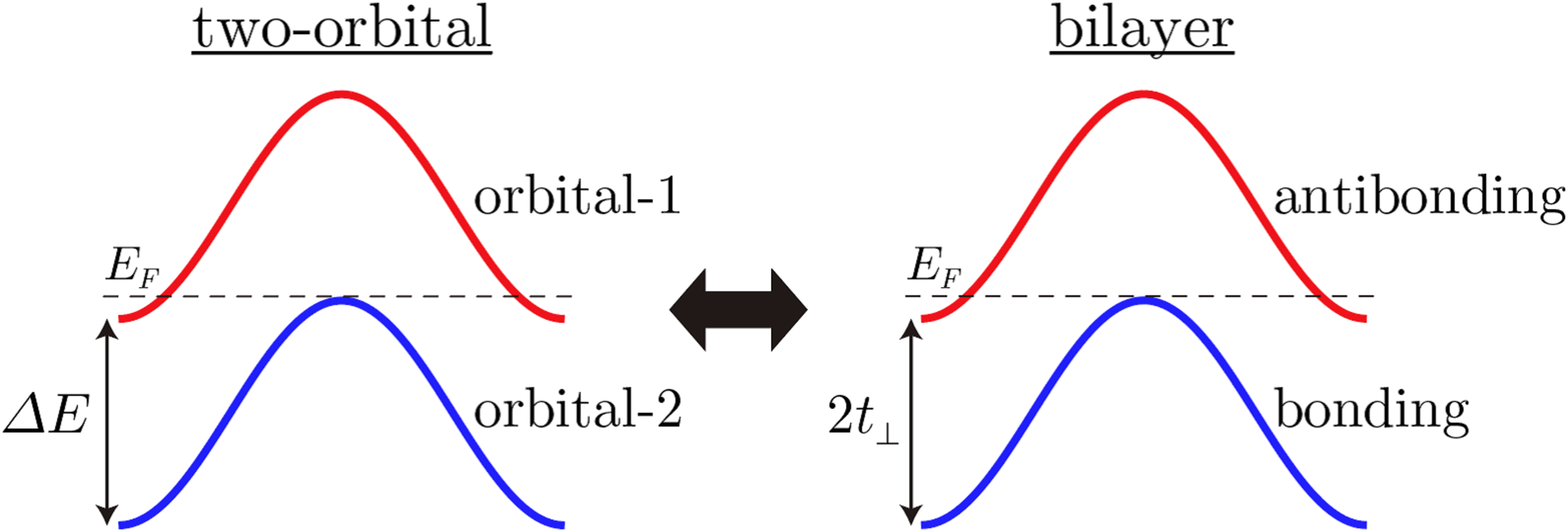}
	\caption{A schematic image of the equivalence between the two-orbital model and the bilayer model.}
	\label{fig1}
\end{figure}

For Ba$_2$CuO$_{3+\delta}$, various theoretical studies have been performed\cite{MaierBa2CuO3,TXiang,LeJHu,LiLiu,NiZou,WangZhang,Yamazaki}, many of which treat multiorbital models. In Ref.\cite{Yamazaki}, two of the present authors have introduced a two-orbital model consisting of Wannier orbitals with $d_{x^2-y^2}$ and $d_{3z^2-r^2}$ symmetries, where $s$-wave superconductivity with a reversed gap sign  between the two bands ($s\pm$-wave pairing) is found to be strongly enhanced near half filling when the energy level offset $\Delta E$ between $d_{3z^2-r^2}$ and $d_{x^2-y^2}$ orbitals is enlarged so that the bottom of the $d_{3z^2-r^2}$ band is positioned just above $E_F$. The strong enhancement of superconductivity was explained by transforming the two-orbital model to the bilayer Hubbard model, where the two orbitals of the former model correspond to the bonding and antibonding orbitals of the latter, and hence the orbital level offset $\Delta E$ in the former transforms to $2t_\perp$ in the latter, with $t_\perp$ being the vertical hopping between the layers in the bilayer model, as schematically depicted in Fig.\ref{fig1}. A more detailed explanation of this transformation is given in the supplemental material\cite{Supplemental}. The bilayer Hubbard model has been intensively studied in the past\cite{Bulut,KA,Scalettar,Hanke,Santos,Mazin,Kancharla,Bouadim,Fabrizio,Zhai,Maier,MaierScalapino,Nakata,MaierScalapino2,Matsumoto2,DKato,Kainth}, and $s\pm$-wave superconductivity is found to be strongly enhanced near half filling when $t_\perp$ is several times larger than the in-plane hopping and $E_F$ lies in the vicinity of one of the bands\cite{Bulut,KA,Maier,MaierScalapino,Nakata,MaierScalapino2,Matsumoto2,DKato}. Nowadays, a band sitting just below (or above) $E_F$ is often referred to as an incipient band, and has attracted interest in the study of iron-based superconductors\cite{DHLee,Hirschfeld,Hirschfeldrev,YBang,YBang2,YBang3,Borisenko,Ding}, bilayer and ladder-type lattices\cite{Kuroki,MaierScalapino2,Matsumoto,Ogura,OguraDthesis,Matsumoto2,DKato,Sakamoto,Kainth}, and flat-band superconductivity\cite{KobayashiAoki,Misumi,Sayyad,Aokireview}. The two-orbital to bilayer transformation is mathematically exact when there is no hybridization between the two orbitals and also $U=U'=J=J'$ is satisfied, where $U$, $U'$, $J$, and $J'$ are the intraorbital repulsion, interorbital repulsion, Hund's coupling, and the pair hopping interaction, respectively\cite{Shinaoka,Yamazaki}. In reality, $U>U'> J, J'$ and the interorbital hybridization is present, but the analogy between the two models turns out to be approximately valid even in the realistic situation\cite{Yamazaki}. 

Actually, a situation where $E_F$ lies close to the $d_{3z^2-r^2}$ band edge is realized in a cuprate (La,Sr)$_2$CuO$_4$. One of the present authors and his colleagues have pointed out that the presence of the $d_{3z^2-r^2}$ band around $E_F$ is the origin of the {\it suppression} of $T_c$ of the $d$-wave superconductivity in this material\cite{Sakakibara,Sakakibara2,Sakakibara3,Sakakibara4} due to the orbital component mixture around the antinodal regime. The difference between (La,Sr)$_2$CuO$_4$ and Ba$_2$CuO$_{3+\delta}$ in Ref.\cite{Yamazaki} is that in the former, the $e_g$ bands are close to 3/4 filling on average (nearly two electrons in $d_{3z^2-r^2}$ and one electron in $d_{x^2-y^2}$), while the latter is closer to half filling.

The above consideration brings us to an idea of realizing incipient-band-enhanced superconductivity in {\it nickelates}, where the $e_g$ bands become half filled on average for the natural Ni$^{2+}$ valence, namely, the $d^8$ electron configuration. Actually, if all the bands other than $d_{x^2-y^2}$ sink below $E_F$ for the $d^8$ configuration, a pair of $d_{x^2-y^2}$ and any other $d$ orbital is half filled (two electrons per two orbital on average), so if a small amount of electrons is doped into such systems, we may expect enhanced $s\pm$-wave superconductivity. To enlarge the energy level offset between $d_{x^2-y^2}$ and $d_{3z^2-r^2}$ orbitals, we adopt a mixed-anion strategy\cite{KageyamaReview}, namely, we chose halogens or hydrogens instead of oxygens as out-of-plane anions (which corresponds to enhancing $t_\perp$ in the bilayer system), for which we end up with the composition of {\it AE}$_2$NiO$_2${\it X}$_2$ ({\it AE}=Ca,Sr, {\it X}=H, F, Cl, Br, I)\cite{commentCano}.

We note that similar materials have been proposed in Ref.\cite{Hirayama} in the context of a recently discovered superconductor (Nd,Sr)NiO$_2$\cite{Hwang}, but there, the aim was to construct ideal single-band systems with a $d^9$ electron configuration, where bands other than Ni $d_{x^2-y^2}$ do not intersect $E_F$ for the mother compound. In fact, here we later conversely discuss a possible relevance of the present $d^8$ proposal to the superconductivity of (Nd,Sr)NiO$_2$.
\begin{figure}
	\includegraphics[width=8.5cm]{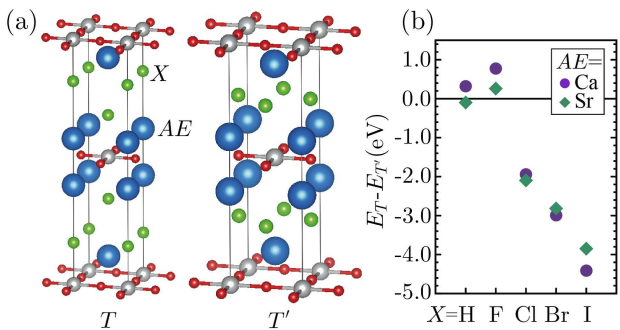}
	\caption{(a) $T$ and $T'$ crystal structures. (b) The total energy difference per formula unit between $T$ and $T'$ structures for various choices of elements.}
	\label{fig2}
\end{figure}

We consider $T$(K$_2$NiO$_4$) and $T'$ (Fig.\ref{fig2}) structures as candidates for possible crystal structures\cite{comment2}. We also take La$_2$NiO$_4$ ($T$ structure) as a reference with a $d^8$ configuration. 
We perform a structural optimization adopting the PBE-GGA exchange-correlation functional\cite{PBE-GGA} and the projector augmented-wave method\cite{Kresse}. We
use the Vienna ab initio Simulation Package (VASP)\cite{VASP1,VASP2,VASP3,VASP4}. A $12\times 12\times 12$ $k$ mesh and a plane-wave cutoff energy of 550 eV were used. 
After the structural optimization, we perform a first-principles band-structure calculation using the WIEN2k code\cite{Wien2k}. We adopt RKmax=7 (6 for oxy-hydrides), and take $12\times 12\times 12$ $k$ mesh in the self-consistent-field calculations. From the calculated band structures, we extract the Wannier functions\cite{Marzari,Souza} of five Ni $3d$ orbitals using the Wien2Wannier\cite{w2w} and Wannier90\cite{Wannier90} codes. Throughout the study, the spin-orbit coupling is neglected.

 In Fig.\ref{fig2}(b), we plot the total energy difference between the two structures for all choices of {\it AE} and {\it X} elements. It is found that for {\it X}=Cl,Br and I, the $T$ structure has lower energy ($E_T-E_{T'}<0$), while the difference between the two is small for {\it X}=H,F. In fact, the phonon calculation of Ba$_{0.5}$La$_{0.5}$NiO$_2$F$_2$ in Ref.\cite{Hirayama} shows the appearance of imaginary modes for the $T$ structure, but not for $T'$, which suggests stability of the latter. The optimized lattice constants presented in the supplemental material exhibit a trend where the in-plane lattice constant $a$ becomes larger as the ion radius of element {\it X} or {\it AE} is increased\cite{Supplemental}.

\begin{figure}
	\includegraphics[width=8.5cm]{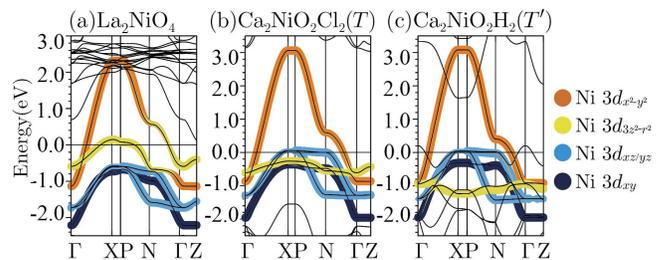}
	\caption{The band structures of (a) La$_2$NiO$_4$, (b) Ca$_2$NiO$_2$Cl$_2$ ($T$), and (c) Ca$_2$NiO$_2$H$_2$ ($T'$). The band dispersion of the five orbital model is superposed to the first principles bands.}
	\label{fig3}
\end{figure}

The first-principles band structures obtained for the optimized lattice structures of La$_2$NiO$_4$ ($T$), Ca$_2$NiO$_2$Cl$_2$ ($T$), and  Ca$_2$NiO$_2$H$_2$ ($T'$) are shown in Fig.\ref{fig3}. The energy dispersion of the five-orbital model is superposed to the first-principles bands for each material. The band structures of materials with other elements are presented in the supplemental material\cite{Supplemental}. It can be noticed that in the mixed-anion materials, the $d_{3z^2-r^2}$ band totally sinks below $E_F$ despite the $d^8$ electron configuration with the $d_{x^2-y^2}$ band partially filled, in sharp contrast to the case of La$_2$NiO$_4$. The entire $d_{xy}$ band also lies below $E_F$, while the top of the $d_{xz/yz}$ bands intersects $E_F$. 

\begin{figure}
	\includegraphics[width=9cm]{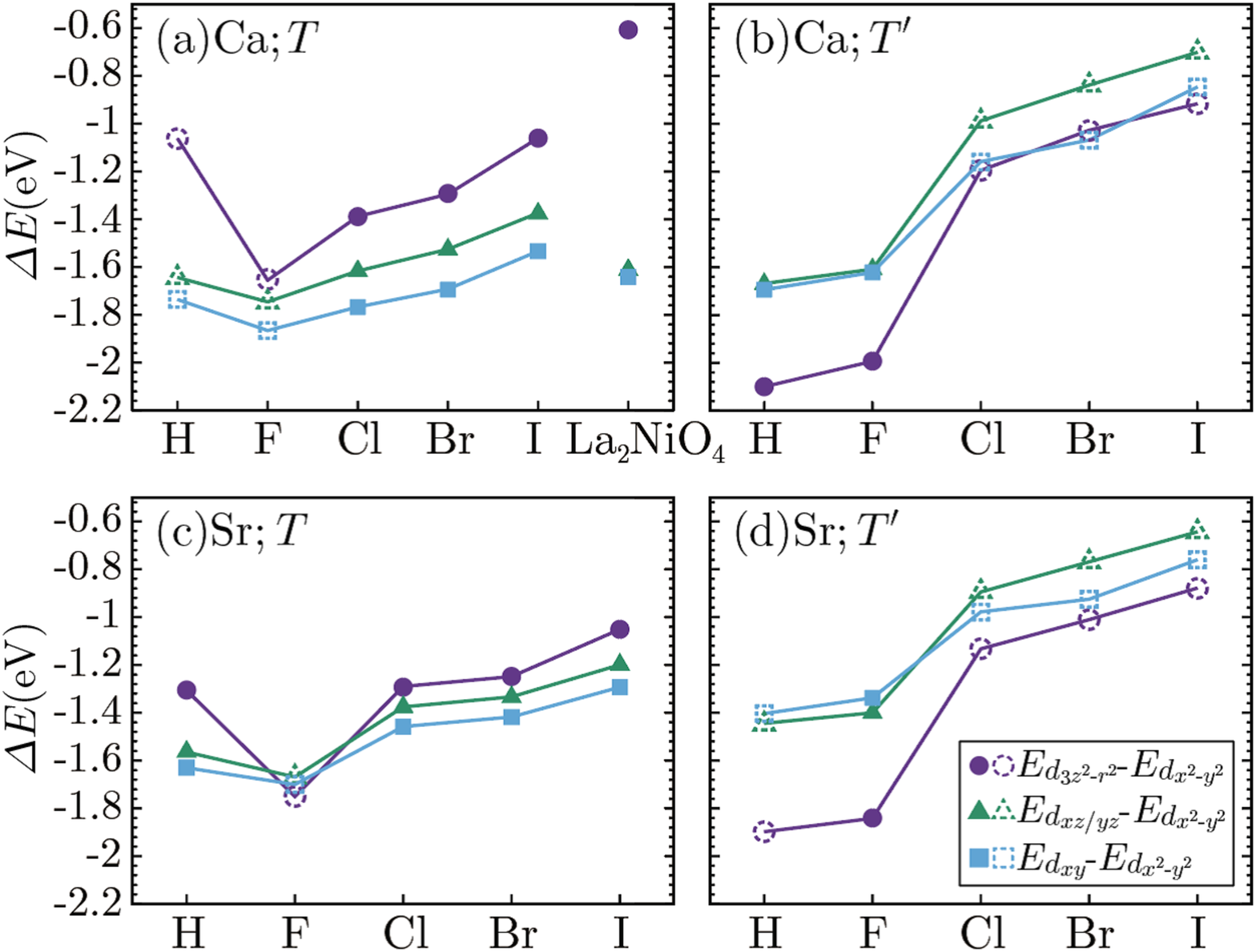}
	\caption{The on-site energy of the $d$ orbitals other than the $d_{x^2-y^2}$ orbital, measured from that of $d_{x^2-y^2}$. (a) {\it AE}=Ca; $T$, (b) Ca; $T'$, (c) Sr; $T$, and (d) Sr; $T'$. Here, the solid (dashed) symbols denote the lattice structure with lower (higher) total energy. As a reference, the values of $\Delta E$ for La$_2$NiO$_4$ are plotted in (a).}
	\label{fig4}
\end{figure}

In Fig.\ref{fig4}, we plot the on-site energy $(\Delta E)$ of the $d$ orbitals with respect to that of $d_{x^2-y^2}$ in the five-orbital model. Let us focus on the crystal structure having the lower total energy (denoted by the solid symbols). $|\Delta E|$ tends to be larger (i.e., the energy level is lowered since $\Delta E<0$) as the ion radius of {\it X} is reduced within {\it X=} Cl, Br and I. This is due to the smaller in-plane lattice constant (shorter Ni-O distance), which pushes up the $d_{x^2-y^2}$ energy level. For the $T'$ structure, the on-site energy of the $d_{3z^2-r^2}$ orbital is strongly reduced due to the absence of the apical anions, but the $t_{2g}$ orbitals are pushed up compared to those in the $T$ structure. If we compare {\it AE}=Sr and Ca, $|\Delta E|$ of the $t_{2g}$ orbitals are reduced in the former compared to the latter, while that of $d_{3z^2-r^2}$ is less affected. This is because the lattice constants $a$ and $c$ are shorter for {\it AE}=Ca than for Sr.

We now analyze superconductivity based on the obtained five-orbital models. 
We assume on-site intra- and interorbital interactions, $U$, $U'$, $J$ and $J'$, and the many-body study of this model is performed within the fluctuation exchange approximation (FLEX)\cite{Bickers}. We mainly adopt $U=4$eV, $J=J'=U/8$, $U'=U-2J$, but also discuss the interaction dependence in the supplemental material\cite{Supplemental}. A relatively large $U$ is taken in accord with a study for LaNiO$_2$\cite{SakakibaraNi}. We obtain the renormalized Green's function by solving the Dyson's equation in a self-consistent calculation.  
The obtained Green's function and the pairing interaction mediated mainly by spin fluctuations are plugged into the linearized Eliashberg equation. 
Since the eigenvalue $\lambda$ of the equation reaches unity at $T=T_c$, here we adopt $\lambda$, obtained at a fixed temperature of $T=0.01$eV, to measure how close the system is to superconductivity. The eigenfunction of the Eliashberg equation will be called the gap function. 
In the FLEX calculation, a $16\times 16\times 4$ $(k_x,k_y,k_z)$ mesh and 2048 Matsubara frequencies were taken. 
\begin{figure}[b]
	\includegraphics[width=8.5cm]{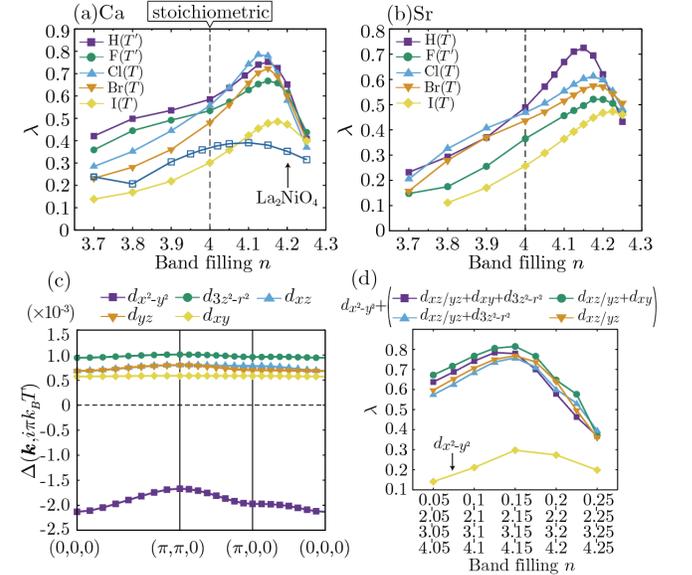}
	\caption{Eigenvalue of the Eliashberg equation $\lambda$ at $T=0.01$eV against the band filling for (a) {\it AE}=Ca and (b) Sr. In (a), also the results for La$_2$NiO$_4$ are plotted as a reference. (c) Gap function of Ca$_2$NiO$_2$Cl$_2$ for $n=4.125$ in the orbital representation. (d) Similar plot of $\lambda$ obtained for various one-, three-, and four-orbital models (see text) of Ca$_2$NiO$_2$Cl$_2$. The horizontal axis is the band filling (for the single-orbital to the five-orbital model from top to bottom).}
	\label{fig5}
\end{figure}

In Figs.\ref{fig5}(a) and \ref{fig5}(b), we plot the eigenvalue of the Eliashberg equation against the band filling, varied assuming a rigid band, for various materials with the crystal structure ($T$ or $T'$) having a lower total energy. In all mixed-anion compounds, $\lambda$ is peaked at around $n=4.1$ to 4.2, which corresponds to 10 to 20\% electron doping starting from the stoichiometric band filling of $n=4$. The compounds having the largest maximum $\lambda$ are Ca$_2$NiO$_2$Cl$_2$ and Ca$_2$NiO$_2$H$_2$. These maximum values of $\lambda$ are much larger compared to that of La$_2$NiO$_4$\cite{comment3}, and they are in fact even larger than that of HgBa$_2$CuO$_4$, a $T_c=$100K superconductor, obtained in the same manner\cite{SakakibaraNi}.  It is also worth mentioning that the Stoner factor for magnetic ordering remains far from unity for the situation when $\lambda$ of superconductivity is optimized, which is typical for the incipient-band-enhanced superconductivity (see supplemental material\cite{Supplemental}).

The gap function presented in Fig.\ref{fig5}(c) for the case of Ca$_2$NiO$_2$Cl$_2$ with $n=4.125$ shows that the four bands other than $d_{x^2-y^2}$ have the same sign of the gap, opposite to that of the $d_{x^2-y^2}$ band (see the supplemental material for the contour plot of the gap\cite{Supplemental}). In this sense, this is $s\pm$-wave pairing with the four bands other than $d_{x^2-y^2}$ being incipient, namely, located below $E_F$, with the $d_{xz/yz}$ band top being the closest to $E_F$. In order to see the role played by each orbital in more detail, we extract one, three, or four orbitals from the five-orbital model, and apply FLEX to those models to obtain $\lambda$ in a similar manner, as depicted in Fig.\ref{fig5}(d). The comparison among the models is made at the same electron doping rates, since the Fermi surfaces are the same among the multiorbital models that contain the $d_{xz/yz}$ orbitals. The $d_{x^2-y^2}$ Fermi surfaces of the multiorbital models are also (nearly) the same as that of the single-band model in the relevant band-filling regime, as explained in more detail in the supplemental material\cite{Supplemental}.
 
From Fig.\ref{fig5}(d), it can be seen that $\lambda$ in the $d_{x^2-y^2}+d_{xz/yz}$ model is strongly enhanced compared to the single-orbital model consisting only of $d_{x^2-y^2}$, indicating the important role played by the incipient $d_{xz/yz}$ bands. Adding $d_{xy}$ to the $d_{x^2-y^2}+d_{xz/yz}$ model somewhat enhances $\lambda$, but not drastically. Our interpretation for this is that the interaction between $d_{xy}$  and $d_{x^2-y^2}$ orbitals enhances superconductivity because the signs of the gap are opposite, but the interaction between $d_{xy}$ and $d_{xz/yz}$ gives a negative contribution to superconductivity because of the same sign of the gap. On the other hand, adding the $d_{3z^2-r^2}$ orbital to the $d_{x^2-y^2}+d_{xz/yz}$ model even slightly suppresses $\lambda$. An argument similar to $d_{xy}$ should apply also to $d_{3z^2-r^2}$, but, in addition, we find that the hybridization between $d_{3z^2-r^2}$ and $d_{x^2-y^2}$ (note that the $d_{x^2-y^2}$ orbital is hybridized only with $d_{3z^2-r^2}$) gives a negative contribution to superconductivity, similarly to what has already been known for the cuprates\cite{Sakakibara,Sakakibara2,Sakakibara3,Sakakibara4}. It is also interesting to note that the $d_{3z^2-r^2}$ gap function has the largest value among the incipient bands, despite the negative contribution of $d_{3z^2-r^2}$ to $\lambda$, and also despite the $d_{3z^2-r^2}$ band lying somewhat away from $E_F$ compared to the top of the $d_{xz/yz}$ bands. We believe that this is a consequence of small $|\Delta E|$ for $d_{3z^2-r^2}$ orbital in Ca$_2$NiO$_2$Cl$_2$, which implies that this band lies closest to $E_F$ among the incipient bands as far as the center of gravity of the band is concerned. More detailed analyses on these issues are given in the supplemental material\cite{Supplemental}.

Since $\lambda$ is dominated by the position of the $d_{xz/yz}$ bands, the materials having a larger energy difference between the $d_{x^2-y^2}$ and $d_{xz/yz}$ orbitals, corresponding to larger $t_\perp$ in the bilayer Hubbard model\cite{Yamazaki}, tend to exhibit larger $\lambda$. On the other hand, it should be stressed that the strong reduction of $\lambda$ for La$_2$NiO$_4$ compared to mixed-anion materials mainly comes from the small $|\Delta E|$ for the $d_{3z^2-r^2}$ orbital; the $d_{3z^2-r^2}$ band intersecting $E_F$ gives rise to the strong development of low-energy spin fluctuations, which works destructively against superconductivity, as in the case of the bilayer Hubbard model with small $t_\perp$\cite{MaierScalapino,Nakata,Matsumoto2,DKato}. A further analysis on the role played by each orbital is given in the supplemental material\cite{Supplemental}.

We note that all of our calculations assume a low-spin state, while nickelates with a $d^8$ configuration can take high-spin states when the $d$-level splittings are small. In fact, the reference system in our study La$_2$NiO$_4$ is actually well known to be in the high-spin state, and also Sr$_2$NiO$_2$Cl$_2$, synthesized in Ref.\cite{Tsujimoto}, was found to be in a high-spin state. On the other hand, a hypothetical infinite layer nickelate SrNiO$_2$, without apical anions, was theoretically shown to have a low-spin ground state\cite{Anisimov}. Hence, large $|\Delta E|$ is not only preferable for enhancing $T_c$, but it is also favorable from the viewpoint of realizing a low-spin ground state. In this sense, {\it AE}=Ca rather than Sr and also {\it X}=Cl rather than Br or I ($T$ structure) is likely to be the best choice from the viewpoint of both realizing a low-spin state and enhancing $T_c$ because of the smaller in-plane lattice constant and hence larger $|\Delta E|$. Our result indicates that superconductivity in {\it AE}$_2$NiO$_2${\it X}$_2$ is optimized by electron doping, which may be realized in actual materials by, e.g., partially substituting Ca$^{2+}$ (or Sr$^{2+}$) with La$^{3+}$. In fact, a pure $d^8$ configuration (band filling of $n=4$) may result in an insulating state not predicted within the present calculation. Electron doping should prevent the material from being an insulator, and hence may also stabilize the preferred low-spin state in actual materials. In total, the materials we propose here as candidates for new unconventional superconductors have the composition forms Ca$_{2-x}$La$_x$NiO$_2$Cl$_2$ and Ca$_{2-x}$La$_x$NiO$_2$H$_2$.

\begin{figure}
	\includegraphics[width=8.5cm]{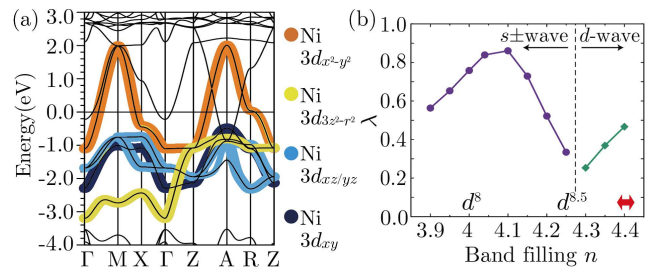}
	\caption{(a) The first principles band structure of LaNiO$_2$ with the five-orbital model superposed. (b) $\lambda$ against the band filling for the model of LaNiO$_2$. $\leftrightarrow$ shows the band filling-regime corresponding to the Sr content where superconductivity is observed in Refs.\cite{Hwang2,Ariando} assuming no residual hydrogens are present.}
	\label{fig6}
\end{figure}

Now, if we increase the amount of electron doping in the present materials, they approach the materials with a $d^9$ electron configuration studied in Refs.\cite{Hirayama,Nomura2} in the context of a recently discovered superconductor (Nd,Sr)NiO$_2$\cite{Hwang}. Then, conversely, it is interesting to see how the $d^9$ and $d^8$ states are connected in (Nd,Sr)NiO$_2$ by hypothetically removing the electrons. Here, we consider LaNiO$_2$ adopting the experimentally determined lattice parameters of NdNiO$_2$ to avoid the ambiguity regarding the treatment of the $f$ electrons in Nd\cite{SakakibaraNi}. Since a previous study has shown that the La $5d$ orbitals have a small effect on spin-fluctuation-mediated superconductivity\cite{SakakibaraNi}, we construct a five Ni-$3d$-orbital model similar to those for {\it AE}$_2$NiO$_2$X$_2$, as shown in Fig.\ref{fig6}(a). The band structure somewhat resembles that of {\it AE}$_2$NiO$_2$Cl$_2$ in that the position of the $d_{3z^2-r^2}$ band is lowered, which is because the apical anions are absent in this infinite layer material, but the difference lies in that the $d_{3z^2-r^2}$ band exhibits strong three dimensionality. We perform a similar FLEX analysis of superconductivity for this model by hypothetically varying the band filling assuming a rigid band. As expected from our results on {\it AE}$_2$NiO$_2${\it X}$_2$, there appears a peak in $\lambda$ close to $n=4$, i.e., the $d^8$ configuration, in addition to the already known $d$-wave superconductivity around $n=4.5$, the $d^9$ configuration. 

Although we believe that the mechanism similar to the cuprates, where only the $d_{x^2-y^2}$ band plays an important role, is a strong candidate for the superconducting mechanism of (Nd,Sr)NiO$_2$\cite{SakakibaraNi,Nomura,Thomale,Kitatani}, the present analysis suggests that their might be an alternative scenario where Ni $3d$ orbitals other than $d_{x^2-y^2}$ play the role of incipient bands. In fact, as studied in Ref.\cite{Held},  (Nd,Sr)NiO$_2$H may form during the reduction process by CaH$_2$, which would reduce the number of $d$ electrons\cite{comment}. Another theoretical study shows that a similar situation can occur also at the SrTiO$_3$ substrate interface\cite{Canointerface}. 
In Fig.\ref{fig6}(b), we indicate the band-filling regime corresponding to the Sr content where superconductivity is observed in Refs.\cite{Hwang2,Ariando} assuming no residual hydrogens are present. This region will be shifted by $p/2$ toward the smaller band-filling regime if residual hydrogens are present in the form (Nd,Sr)NiO$_2$H$_p$.  It will then be an interesting future problem to carefully examine the Ni valence in (Nd,Sr)NiO$_2$. Also, the gap function would be of interest since the $d$ wave should be realized for the $d^9$ pairing mechanism, whereas the $s$ wave is expected for the $d^8$ scenario. In fact, while finalizing the present paper, a tunneling spectroscopy experiment has been reported\cite{HHWen}, where $s$-wave and $d$-wave-like gaps have been observed, depending on the position on the sample. The experiment has been interpreted in terms of the gap with a different symmetry opening on different Fermi surfaces, which was suggested theoretically\cite{Dasgupta}. From our viewpoint, the position dependence of the pairing symmetry might originate from the inhomogeneity of the residual hydrogen remained during the reduction process\cite{FCZhang}.

To conclude, we have designed unconventional nickelate superconductors with a nearly $d^8$ electron configuration, where superconductivity is enhanced by the bands other than $d_{x^2-y^2}$ playing a role of the incipient bands. The key idea is to exploit the mixed-anion strategy to enlarge the on-site energy difference between $d_{x^2-y^2}$ and other $d$ orbitals. This corresponds to increasing the interlayer hopping $t_\perp$ of the bilayer Hubbard model, which is known to exhibit $s\pm$-wave superconductivity strongly enhanced by the incipient-band effect for large $t_\perp$. The possible relevance of the present proposal to the observation of superconductivity in (Nd,Sr)NiO$_2$ is an interesting future problem, where a non-rigid-band variation of the incipient-band dispersion due to residual hydrogens\cite{Held} should be taken into account. 

\begin{acknowledgments}
We acknowledge Kimihiro Yamazaki, Hirofumi Sakakibara, and Hideo Aoki for valuable discussions. Parts of the numerical calculations were performed at the Supercomputer Center, Institute for Solid State Physics, University of Tokyo. This study has been supported by JSPS KAKENHI Grant Numbers JP18H01860, JP19H04697, and JP19H05058.
\end{acknowledgments}


\begin{thebibliography}{99}
\bibitem{comment2} One of the reasons we consider these structures is because they are well known in cuprates. Further estimation of their stability as nickelates is left for future study.
\end{thebibliography}
\end{document}


\preprint{APS/123-QED}

\title{
Supplemental material: Designing nickelate superconductors with $d^8$ configuration exploiting mixed-anion strategy
} 
\author{Naoya Kitamine}
\author{Masayuki Ochi}
\author{Kazuhiko Kuroki}
\affiliation{Department of Physics, Osaka University, 1-1 Machikaneyama-cho, Toyonaka, Osaka, 560-0043, Japan}

\date{\today}

\pacs{ }
\maketitle

\setcounter{equation}{0}
\setcounter{figure}{0}
\setcounter{table}{0}
\setcounter{page}{1}

\renewcommand{\theequation}{S\arabic{equation}}
\renewcommand{\thefigure}{S\arabic{figure}}
\renewcommand{\bibnumfmt}[1]{[S#1]}
\renewcommand{\citenumfont}[1]{S#1}
\renewcommand{\thetable}{S\arabic{table}}

\section{The equivalence between the two-orbital model and the bilayer model}
As explained in the main text, our idea was inspired by a study on a superconducting mechanism of Ba$_2$CuO$_{3+\delta}$\cite{Yamazaki}, where the near equivalence between the two-orbital model and the bilayer model\cite{Shinaoka} plays a key role. Here, we briefly review this equivalence between the two models.

We start with a bilayer Hubbard model, where the two-layers are connected by interlayer hoppings, including the vertical hoppings $t_\perp$ within unit cells.
We label the two layers by $i=1,2$, and the unit cells are labeled by $m$.
The electron-electron interaction part of the Hamiltonian of the bilayer Hubbard model, in standard notations, is given as,

\begin{eqnarray}
H_{\mathrm{int}}^{\mathrm{bilayer}}=U\sum_{m}\sum_{i=1,2}n_{mi\uparrow}n_{mi\downarrow}  \label{int}.
\end{eqnarray}

Let us consider the following transformation $R$, which transforms the above layer representation ($c$) to the orbital representation ($d$) consisting of bonding (denoted as $b$) and antibonding ($a$) orbitals, 

\begin{eqnarray}
\left(
\begin{array}{c}
d_{ma\sigma} \\
d_{mb\sigma} 
\end{array}
\right)
=
R \left(
\begin{array}{c}
c_{m1\sigma} \\
c_{m2\sigma} 
\end{array}
\right)
=
\frac{1}{\sqrt{2}} \left(
\begin{array}{cc}
1 & 1 \\
-1 & 1 
\end{array}
\right)
\left(
\begin{array}{c}
c_{m1\sigma} \\
c_{m2\sigma} 
\end{array}
\right). \nonumber \\
\end{eqnarray}
Applying this transformation, (\ref{int}) becomes 
\begin{eqnarray}
&& \frac{U}{2}\sum_{m,a}n_{ma\uparrow}n_{ma\downarrow}+\frac{U}{2}\sum_{m,a\neq b}n_{ma\uparrow}n_{mb\downarrow} \nonumber \\
&&-\frac{U}{2}\sum_{m,a\neq b}d_{ma\uparrow}^\dag d_{ma\downarrow}d_{mb\downarrow}^\dag d_{mb\uparrow} \\
&&+\frac{U}{2}\sum_{m,a\neq b}d_{ma\uparrow}^\dag d_{ma\downarrow}^\dag d_{mb\downarrow}d_{mb\uparrow}. \nonumber 
\end{eqnarray}
This is exactly the same as the on-site electron-electron interaction Hamiltonian of a two-orbital model, where the intra-orbital, inter-orbital, Hund's coupling and pair-hopping interactions are all equal to $U/2$.

As for the kinetic energy part of the bilayer model, by Fourier transforming $c_{mi\sigma}^\dag$, $c_{mi\sigma}$ to $c_{{\bm k}i\sigma}^\dag$, $c_{{\bm k}i\sigma}$, respectively, it can be expressed as,  

\begin{eqnarray}
H_{\mathrm{kin}}^{\mathrm{bilayer}}&=&
\left(
\begin{array}{cc}
c_{{\bm k}1\sigma}^\dag & c_{{\bm k}2\sigma}^\dag
\end{array}
\right)
\hat{H}({\bm k)}
\left(
\begin{array}{c}
c_{{\bm k}1\sigma}  \\
c_{{\bm k}2\sigma}
\end{array}
\right) \nonumber \\
&=&
\left(
\begin{array}{cc}
c_{{\bm k}1\sigma}^\dag & c_{{\bm k}2\sigma}^\dag
\end{array}
\right)
\left(
\begin{array}{cc}
\varepsilon_1({\bm k}) & t_\perp+\varepsilon^\prime({\bm k}) \\
t_\perp+\varepsilon^\prime({\bm k}) & \varepsilon_2({\bm k}) 
\end{array}
\right)
\left(
\begin{array}{c}
c_{{\bm k}1\sigma}  \\
c_{{\bm k}2\sigma}
\end{array}
\right). \nonumber  \\
\end{eqnarray}
Here, the diagonal terms correspond to the kinetic energy originating from the intralayer electron hoppings, and the off-diagonal terms come from the interlayer hoppings, where we have separated the vertical hopping $t_\perp$, a constant, and $\varepsilon^\prime({\bm k})$, i.e., the contribution coming from other (diagonal) interlayer hoppings, which exhibits wave number dependence.  Applying the transformation $R$, this becomes, 

\begin{eqnarray}
&&R\hat{H}({\bm k})R^\dag=  \nonumber \\
&&\left(
\small
\begin{array}{cc}
\frac{1}{2}\left(\varepsilon_1({\bm k})+\varepsilon_2({\bm k})\right)+\varepsilon^\prime({\bm k})+t_\perp & \hspace{-3mm}\frac{1}{2}\left(\varepsilon_2({\bm k})-\varepsilon_1({\bm k})\right) \\
\frac{1}{2}\left(\varepsilon_2({\bm k})-\varepsilon_1({\bm k})\right)  & \hspace{-3mm}
\frac{1}{2}\left(\varepsilon_1({\bm k})+\varepsilon_2({\bm k})\right)-\varepsilon^\prime({\bm k})-t_\perp
\end{array}
\right). \nonumber \\
\label{S5}
\end{eqnarray}
Now the diagonal terms are the band dispersion within each orbital, so $2t_\perp$ corresponds to the energy level off-set $\Delta E$ between the bonding and antibonding bands. The off-diagonal terms represent the hybridization between the two-orbitals; if the two layers are not equivalent, namely, $\varepsilon_2({\bm k})\neq \varepsilon_1({\bm k})$, there appears hybridization between the two orbitals obtained by the transformation $R$, which treats the two layers equally. (Namely, $R$ obviously does not diagonalize the kinetic energy part of the bilayer Hamiltonian.)

Conversely, viewing from the two-orbital model side, we can say from the above argument that the two-orbital model with $U=U'=J=J'$ and without hybridization between the orbitals can be transformed to a bilayer Hubbard model with appropriate intra- and interlayer hoppings, where the orbital level offset $\Delta E$ is transformed to $2t_\perp$. In actual materials, $U>U'>J\sim J'$, but the finding in Ref.\cite{Yamazaki} and also in the present study is that the approximate equivalence between the two-orbital and bilayer models holds relatively well even for realistic values of $U$, $U'$, $J$ and $J'$, as far as the strong enhancement of superconductivity for large $\Delta E$ is concerned. 

If orbital hybridization is present in the two-orbital model, this will give rise to an inequality of the two layers in the corresponding bilayer model, as seen from  eqn.(\ref{S5}). The superconducting properties of bilayer models with inequivalent layers are not well known, to our knowledge, but we will see the effect of the orbital hybridization in the last section of this supplemental material.

\section{Lattice constants and band structures}
\begin{figure}[b]
	\includegraphics[width=8cm]{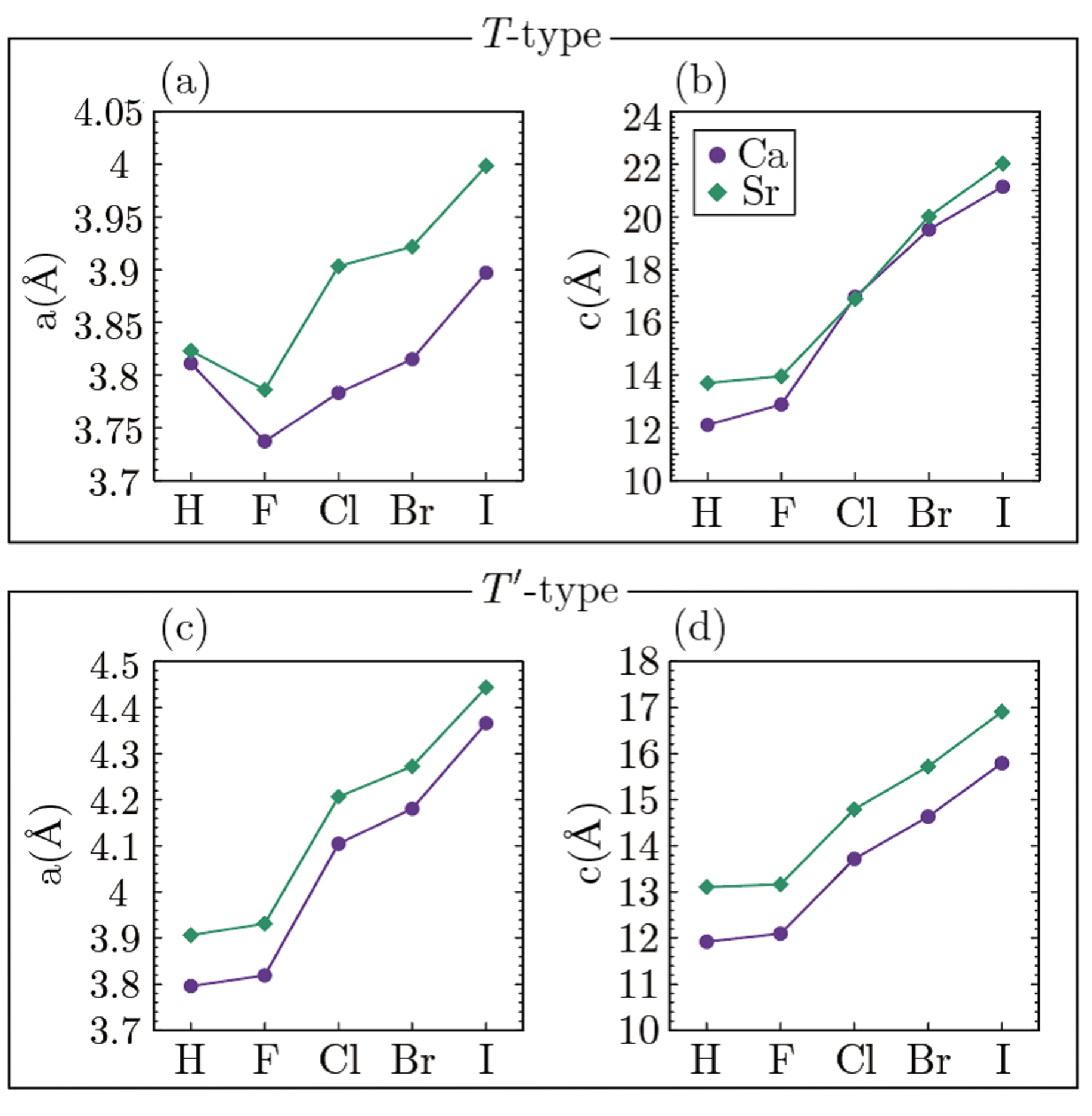}
	\caption{Lattice constants for all the materials considered.}
	\label{figS1}
\end{figure}

\begin{figure}
	\includegraphics[width=6.5cm]{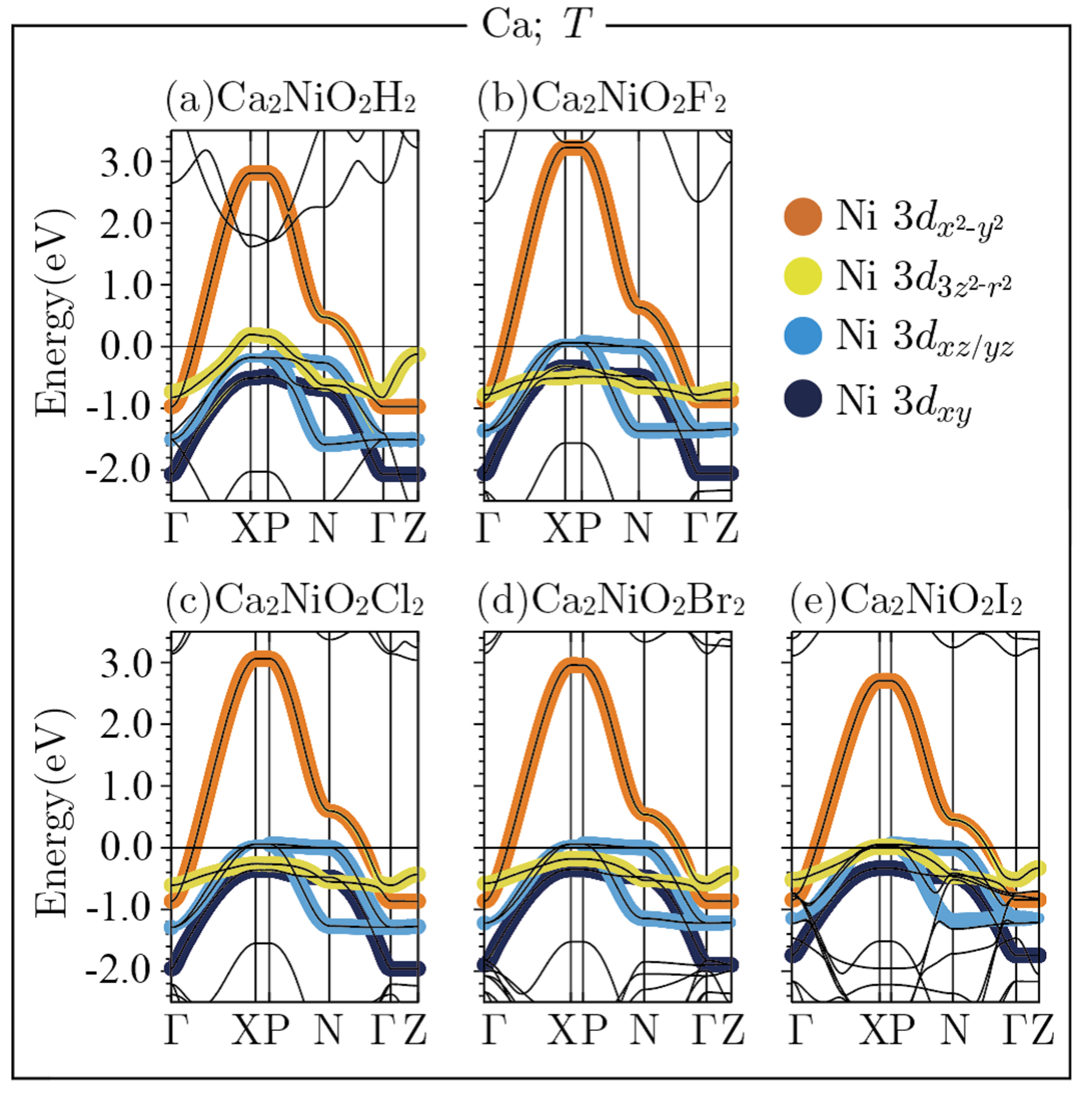}
	\caption{Band structures of materials with {\it AE}=Ca and the $T$ structure. The band dispersion of the five orbital model is superposed to the first principles bands.}
	\label{figS2}
\end{figure}

\begin{figure}
	\includegraphics[width=6.5cm]{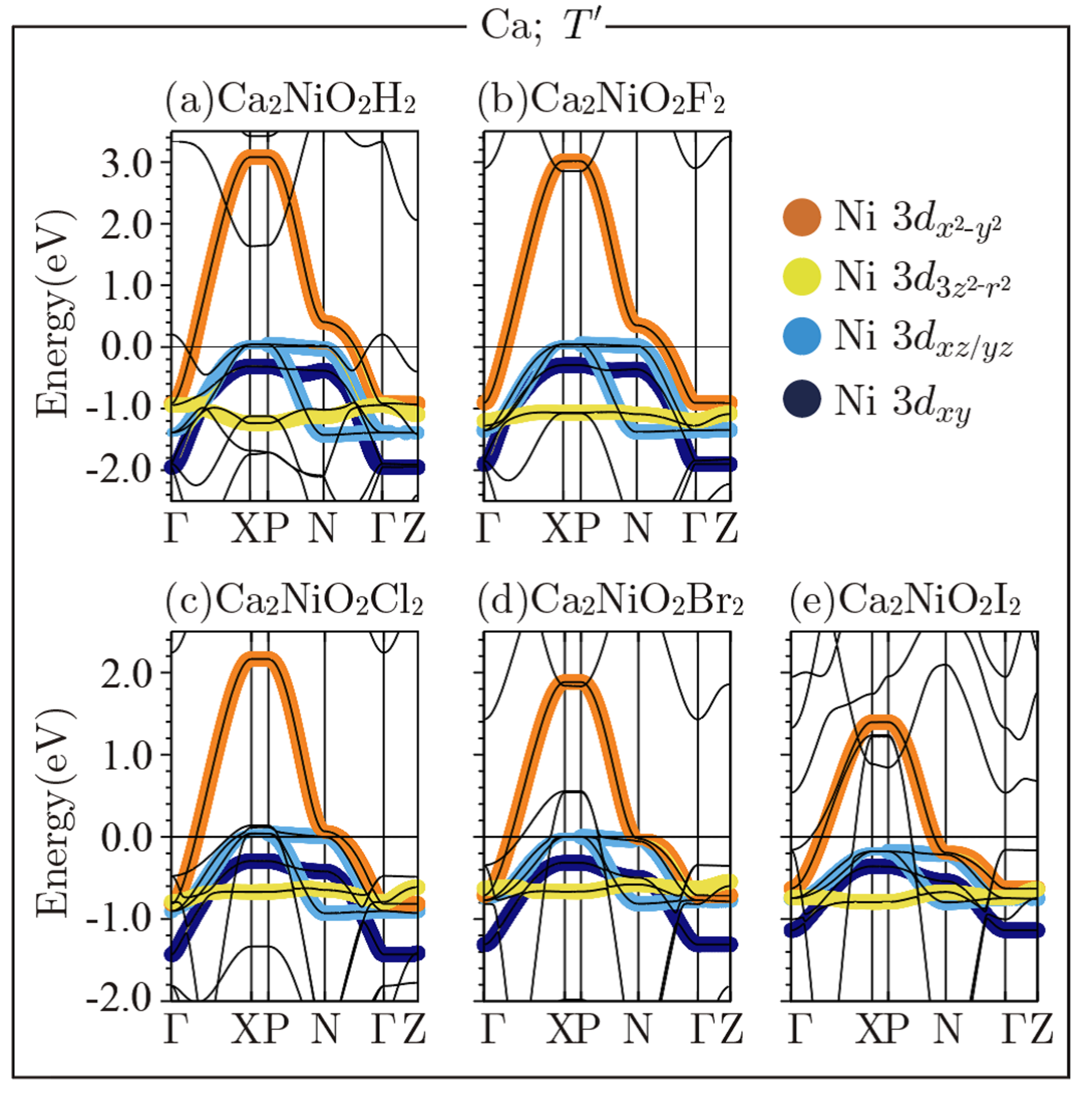}
	\caption{Similar figure as in Fig.\ref{figS2} with {\it AE}=Ca and the $T'$ structure.}
	\label{figS3}
\end{figure}

\begin{figure}
	\includegraphics[width=6.5cm]{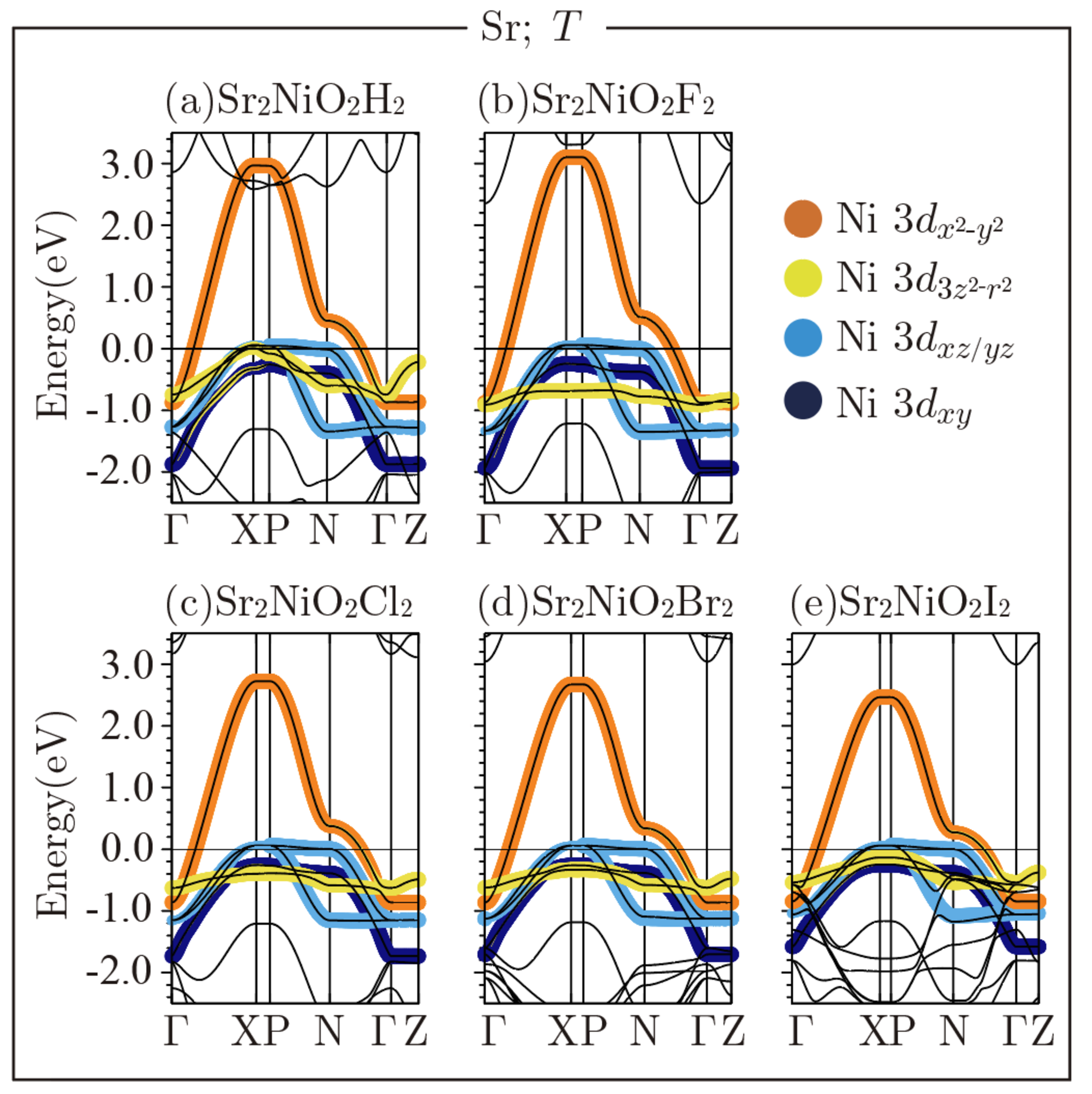}
	\caption{Similar figure as in Fig.\ref{figS2} with {\it AE}=Sr and the $T$ structure.}
	\label{figS4}
\end{figure}

\begin{figure}
	\includegraphics[width=6.5cm]{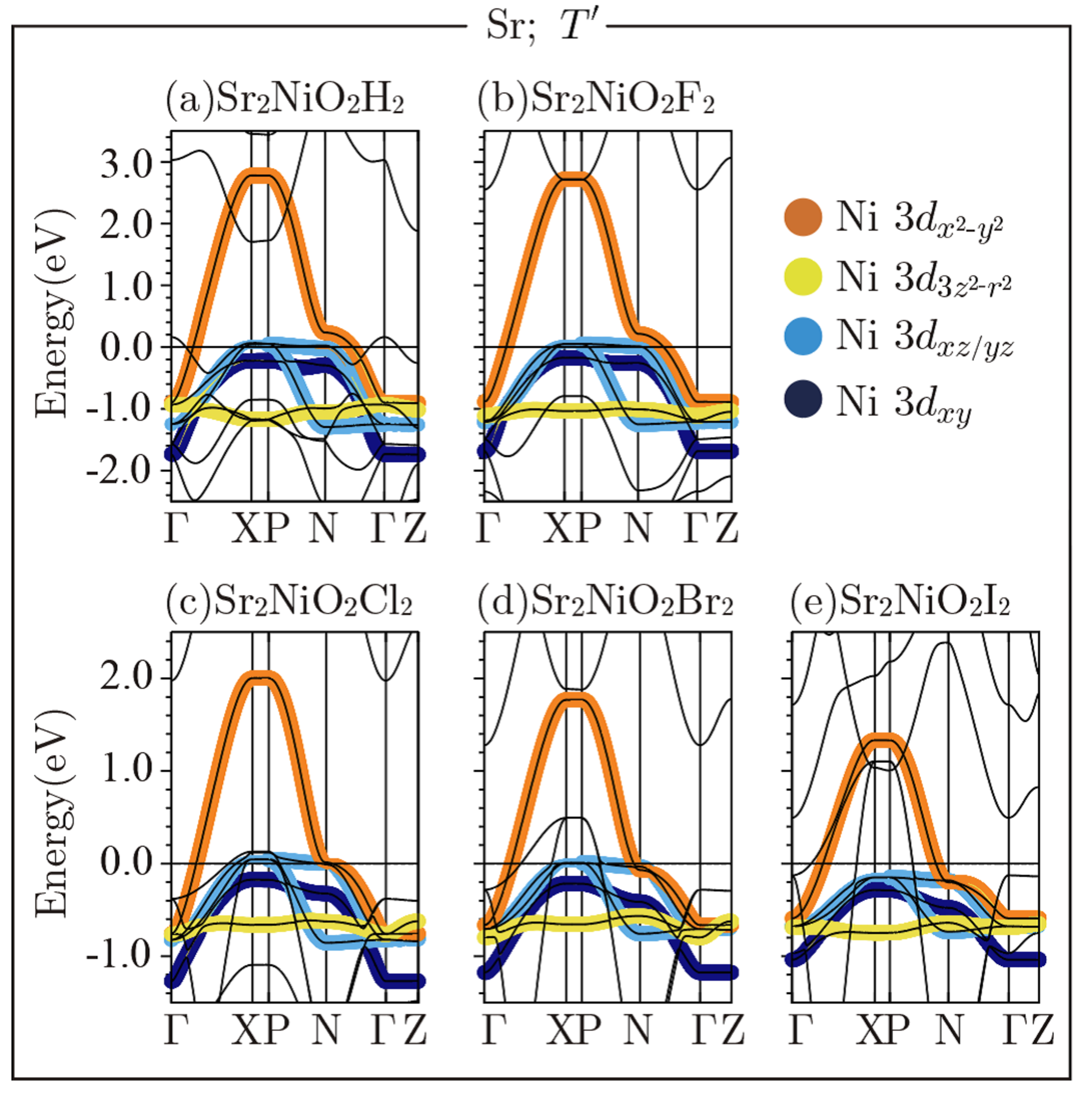}
	\caption{Similar figure as in Fig.\ref{figS2} with {\it AE}=Sr and the $T'$ structure.}
	\label{figS5}
\end{figure}

In Fig.\ref{figS1} and Figs.\ref{figS2}-\ref{figS5}, we plot the lattice constants obtained by structural optimization and the band structures, respectively, for all the materials considered in the present study. The lattice constants are larger for {\it AE} and/or {\it X} with larger ionic radius, except for the case of H in the $T$ structure. As for the band structures, the relative position between the bands reflects the tendency seen in the on-site energy of each orbital (Fig.4 of the main text).

\begin{figure}
	\includegraphics[width=8cm]{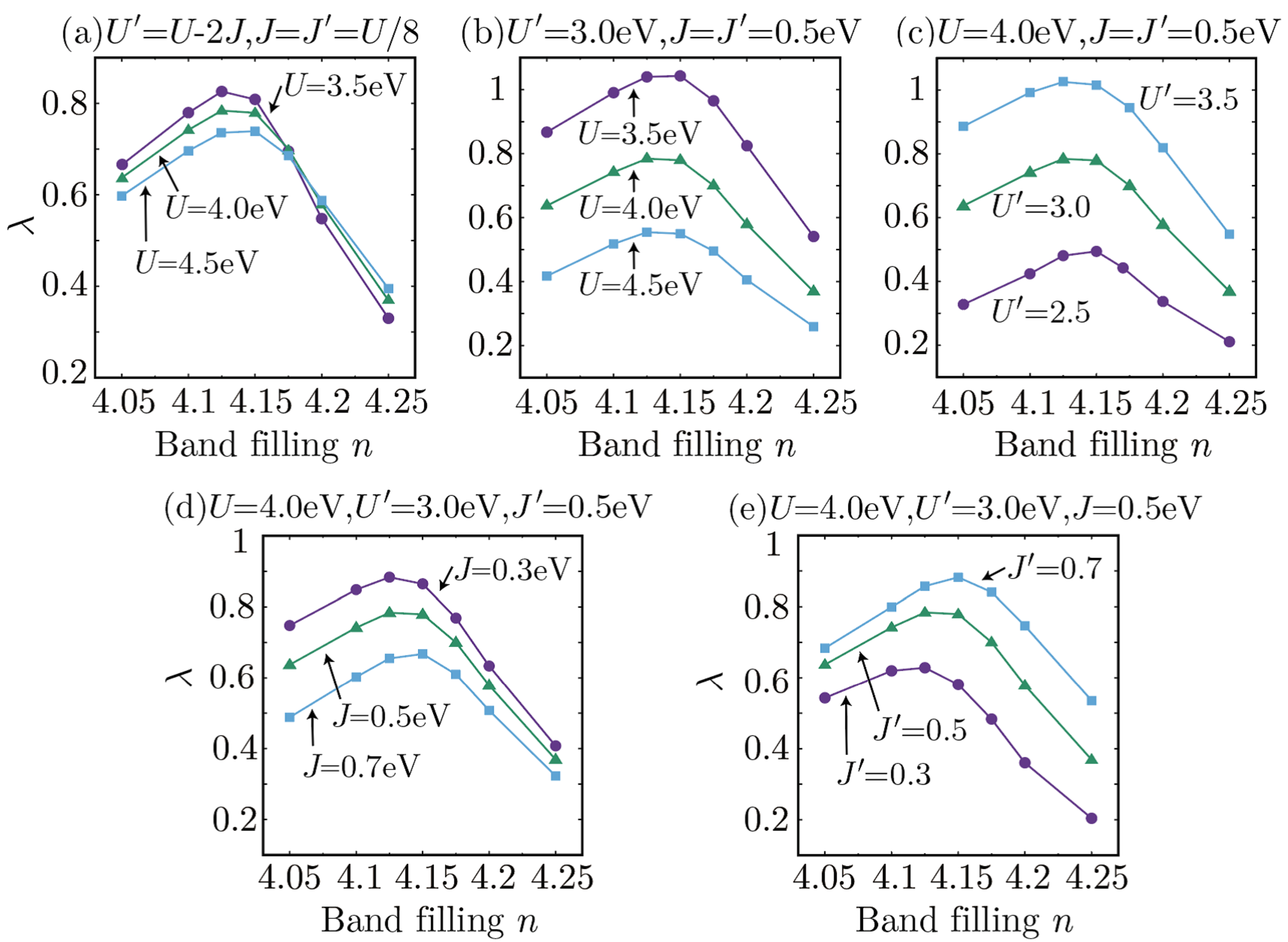}
	\caption{Eigenvalue of the Eliashberg equation $\lambda$ of the five orbital model of Ca$_2$NiO$_2$Cl$_2$ plotted against the band filling $n$ for various sets of interaction parameters. In (a), the interactions are varied maintaining $U'=U-2J$ so as to satisfy the orbital rotational symmetry. From (b) to (d), one of $U$, $U'$, $J$, $J'$ is varied, while the others are fixed.}
	\label{figS6}
\end{figure}
\begin{figure}
	\includegraphics[width=8cm]{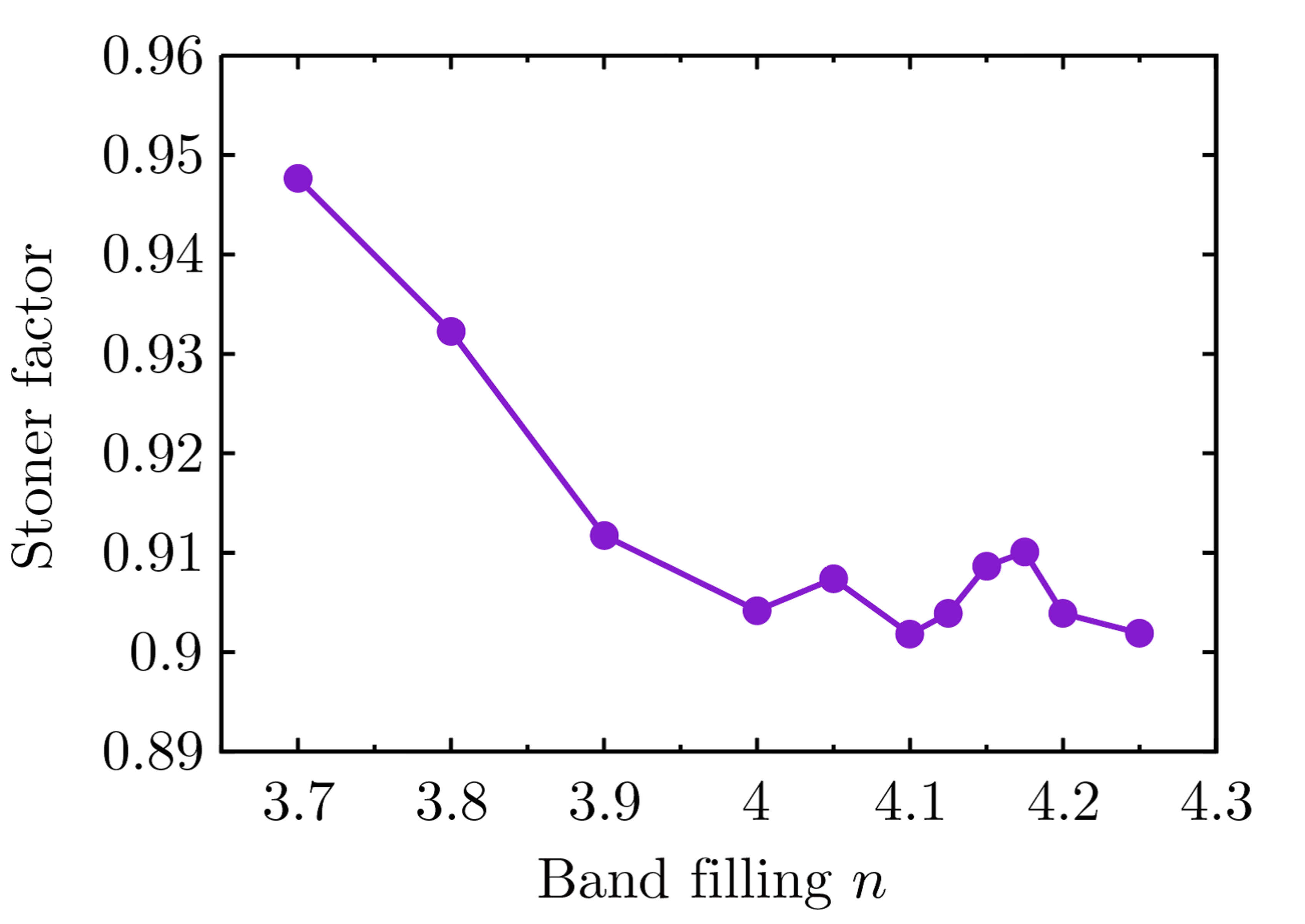}
	\caption{The Stoner factor of the five orbital model of Ca$_2$NiO$_2$Cl$_2$ calculated at $T=0.01$eV within FLEX, plotted against the band filling.}
	\label{figS7}
\end{figure}

\section{Dependence on the interactions}
Here, we vary the interaction values in the five orbital model of Ca$_2$NiO$_2$Cl$_2$ in various manners and see how the eigenvalue of the Eliashberg equation is affected. In Fig.\ref{figS6}(a), the interactions are varied maintaining $U'=U-2J$ so as to satisfy orbital-rotation symmetry. Within the varied parameter range, larger $U$ results in a suppression (slight enhancement) of superconductivity for $n<4.2$ ($n>4.2$). From Fig.\ref{figS6} (b) to (d), one of $U$, $U'$, $J$, $J'$ is varied, while others are fixed. The overall tendency is that increasing $U'$ and $J'$ enhances superconductivity, while increasing $U$ and $J$ degrades it.

\section{Stoner factor for the magnetic ordering}
The tendency toward magnetic ordering can be measured by calculating the Stoner factor within FLEX. In the spin-fluctuation mediated superconductivity, the Stoner factor typically becomes very close to unity (larger than 0.95, where the Stoner factor = 1 signals magnetic ordering) as the eigenvalue of the Eliashberg equation increases, which indicates that superconductivity and antiferromagnetism closely compete with each other\cite{Arita3D}. It is known that this is not the case for the $s\pm$-wave superconductivity enhanced by the incipient band\cite{OguraDthesis}. This is because the incipient-band mechanism exploits finite energy spin fluctuations. In fact, for the five orbital model of the proposed materials, the Stoner factor remains to stay away from unity in the parameter regime where superconductivity is optimized. As a typical example, in Fig.\ref{figS7} we plot the Stoner factor calculated for the five orbital model of Ca$_2$NiO$_2$Cl$_2$ with $U=4$eV, $J=J'=U/8$, $U'=U-2J$, and $T=0.01$eV.
\begin{figure}
	\includegraphics[width=8cm]{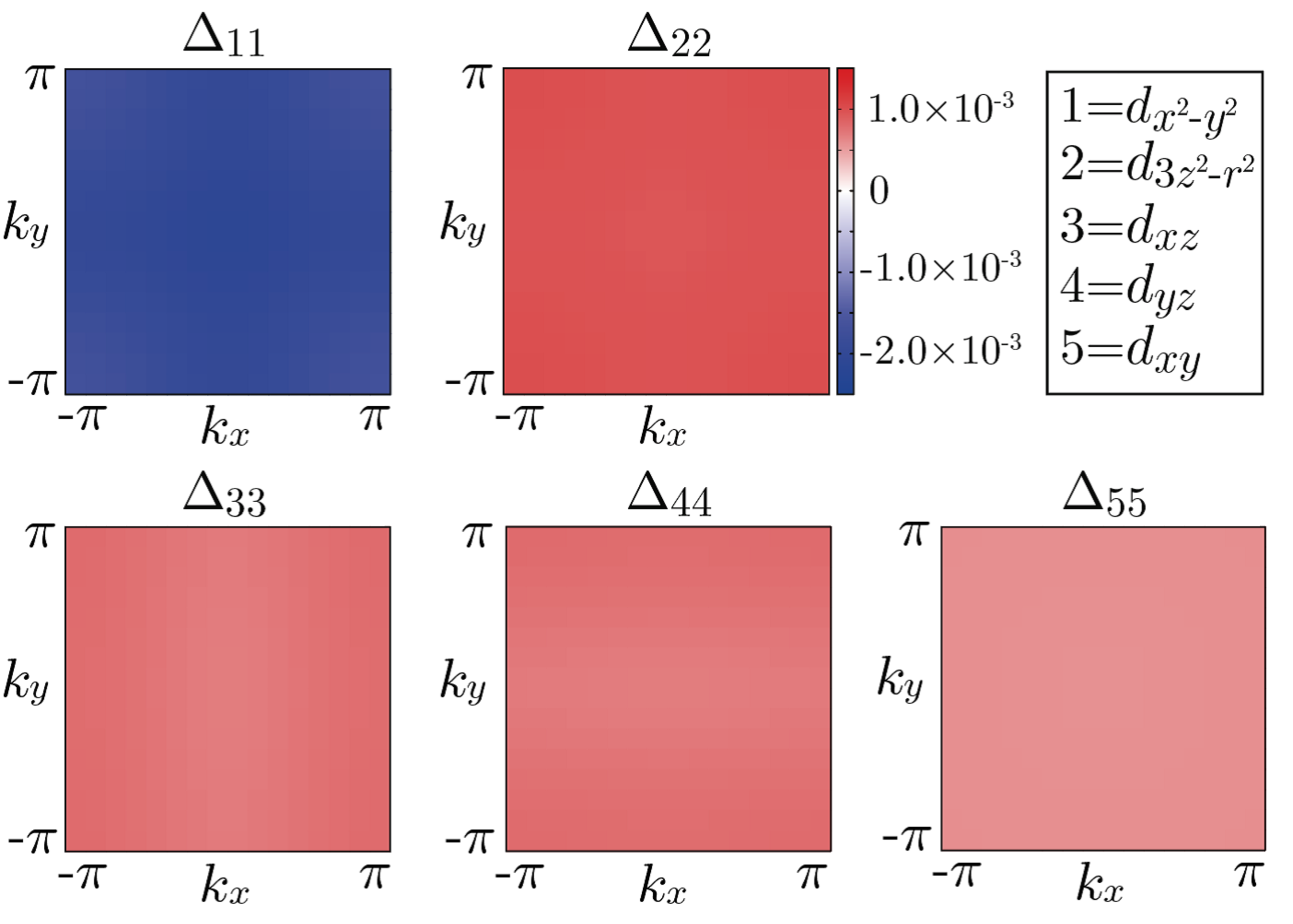}
	\caption{Contour plot of the gap function of the five orbital model of Ca$_2$NiO$_2$Cl$_2$ for $n=4.125$. Only the orbital-diagonal components are shown.}
	\label{figS8}
\end{figure}
\begin{figure}
	\includegraphics[width=8cm]{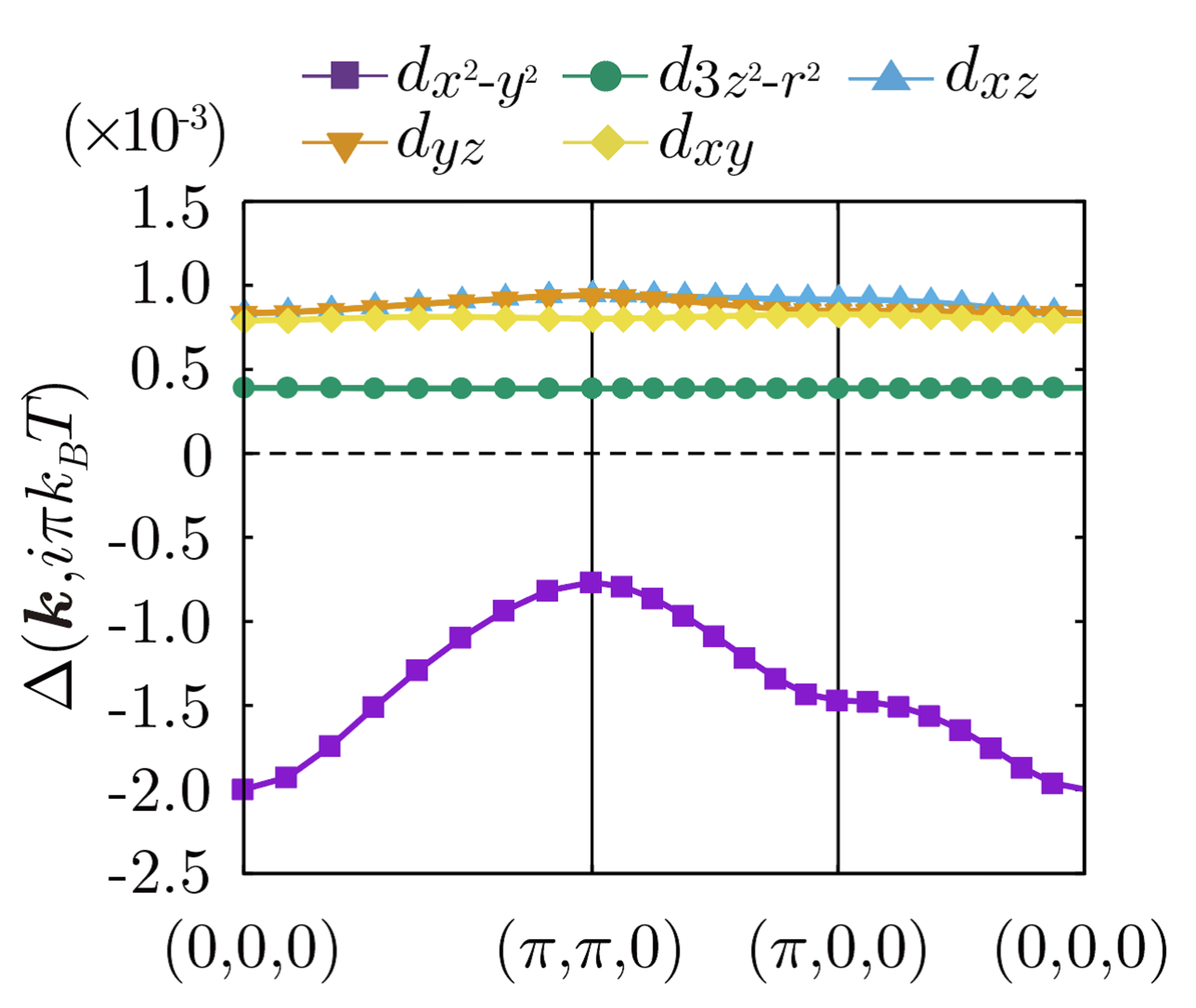}
	\caption{The gap function of the five orbital model of Ca$_2$NiO$_2$H$_2$ for $n=4.15$ in the orbital representation.}
	\label{figS9}
\end{figure}
\begin{figure}
	\includegraphics[width=8cm]{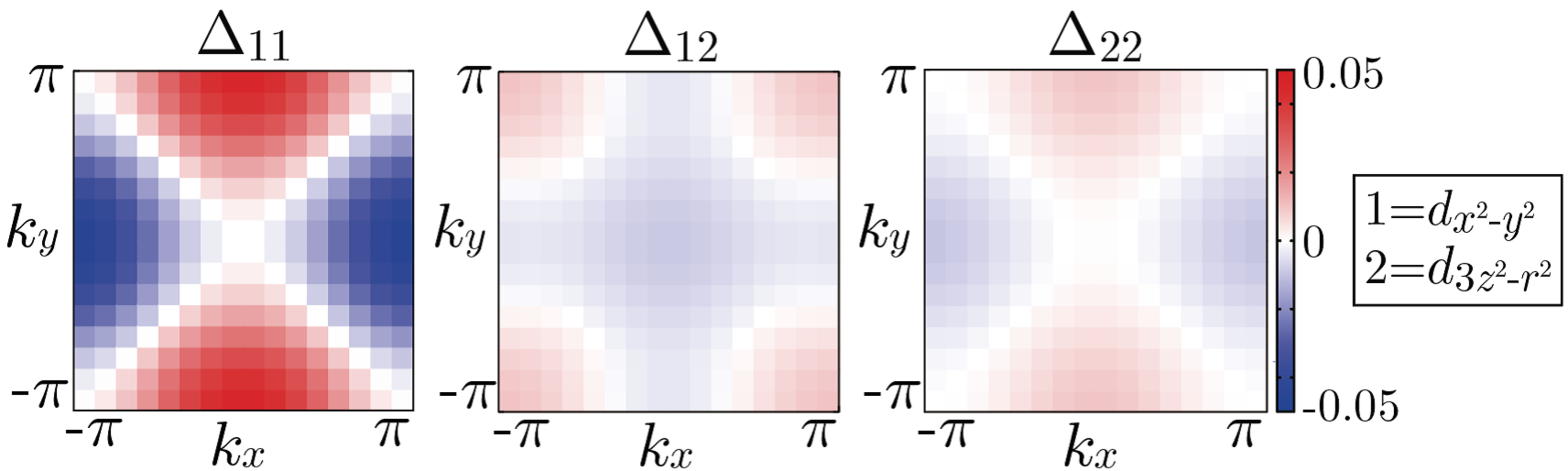}
	\caption{Contour plot of the gap function of the five orbital model of La$_2$NiO$_4$ for $n=4.125$. Here, we plot the off-diagonal component since it has appreciable amplitude.}
	\label{figS10}
\end{figure}

\section{Gap function}
In Fig.\ref{figS8}, we show the contour plot of the gap function for the five orbital model of Ca$_2$NiO$_2$Cl$_2$. As already seen in Fig.5(c) of the main text, this is an $s\pm$-wave pairing with small $k$ dependence of the gap, where the gap changes sign between the $d_{x^2-y^2}$ and incipient $d$ orbitals. As mentioned in the main text, it is interesting to note that the $d_{3z^2-r^2}$ orbital has the largest gap function among the incipient bands despite the fact that the $d_{3z^2-r^2}$ band lies somewhat away from $E_F$ compared to the top of the $d_{xz/yz}$ bands. This may be understood as follows. Since the gap function is barely wave-number dependent, it is likely that not just the position of the top of the band, but also the center of gravity of the incipient bands is important. Since the $d_{3z^2-r^2}$ orbital has the highest on-site energy ($|\Delta E|$ being the smallest) among the incipient orbitals in Ca$_2$NiO$_2$Cl$_2$, the $d_{3z^2-r^2}$ band can be considered as positioned closest to $E_F$ as far as the center of gravity of the band is concerned. In fact, as shown in Fig.\ref{figS9}, in $T'$-type Ca$_2$NiO$_2$H$_2$, where the $d_{3z^2-r^2}$ orbital has the lowest on-site energy among the five orbitals, the $d_{3z^2-r^2}$ gap is found to be the smallest.

For La$_2$NiO$_4$, there is no chance for superconductivity since $\lambda$ is small, and, in the first place, we assume an unrealistic low-spin state. Nonetheless, we plot the inter-orbital and intra-orbital components of the gap function in Fig.\ref{figS10}. It is mainly intra-orbital $d_{x^2-y^2}$-wave within the $d_{x^2-y^2}$ orbital, but there is also appreciable inter-orbital component between $d_{x^2-y^2}$ and $d_{3z^2-r^2}$ orbitals. This is similar to the case of Ba$_2$CuO$_{3+\delta}$ studied in Ref.\cite{Yamazaki}  with small level offset between $d_{x^2-y^2}$ and $d_{3z^2-r^2}$ orbitals.

\section{Orbital decomposition}
\begin{figure}[h]
	\includegraphics[width=8cm]{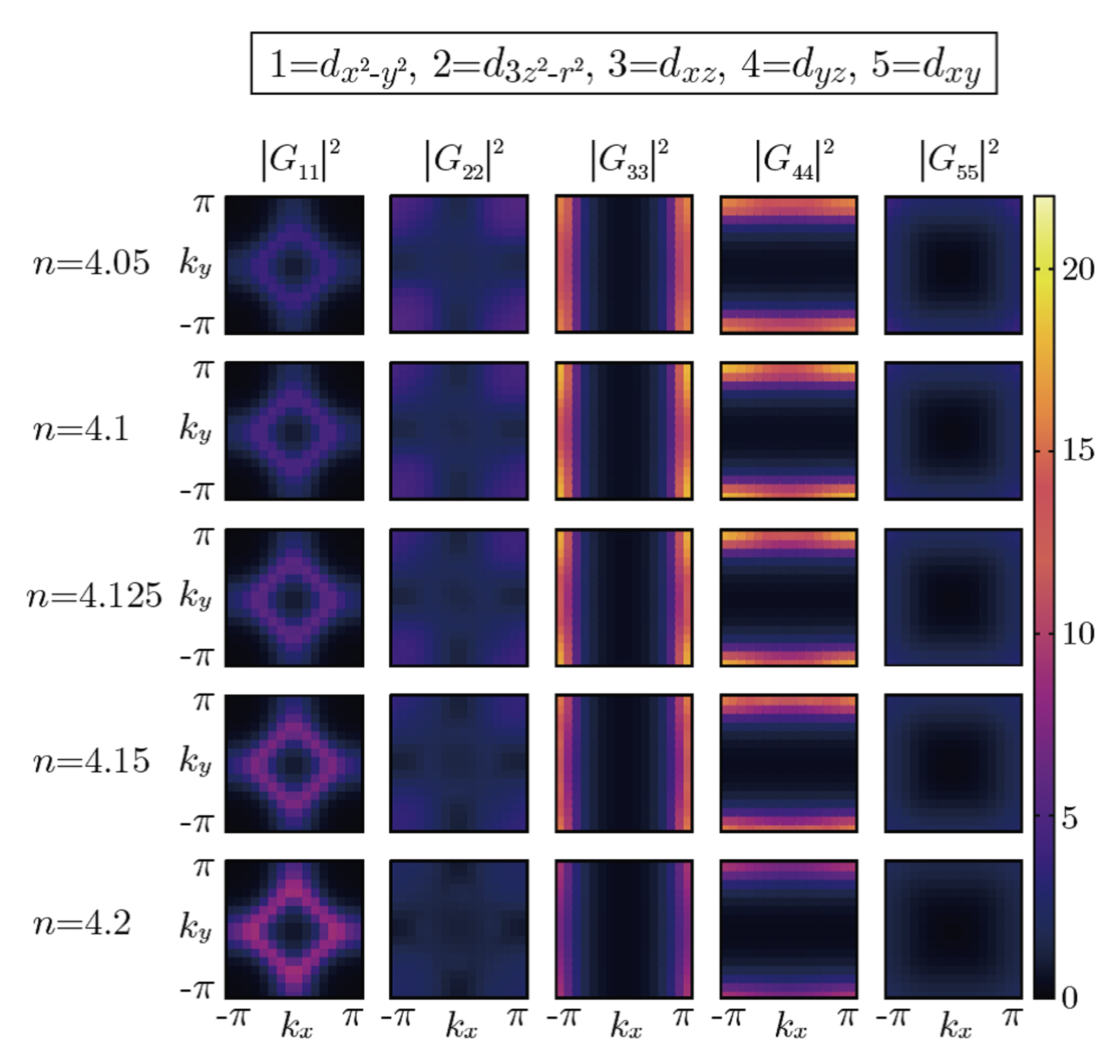}
	\caption{Contour plots of $|G_{ii}({\bm k},i\pi k_BT)|^2$ for the five orbital model of Ca$_2$NiO$_2$Cl$_2$ for $n=4.05$, 4.1, 4.125, 4.15, and 4.2.}
	\label{figS11}
\end{figure}

\begin{figure}[h]
	\includegraphics[width=8cm]{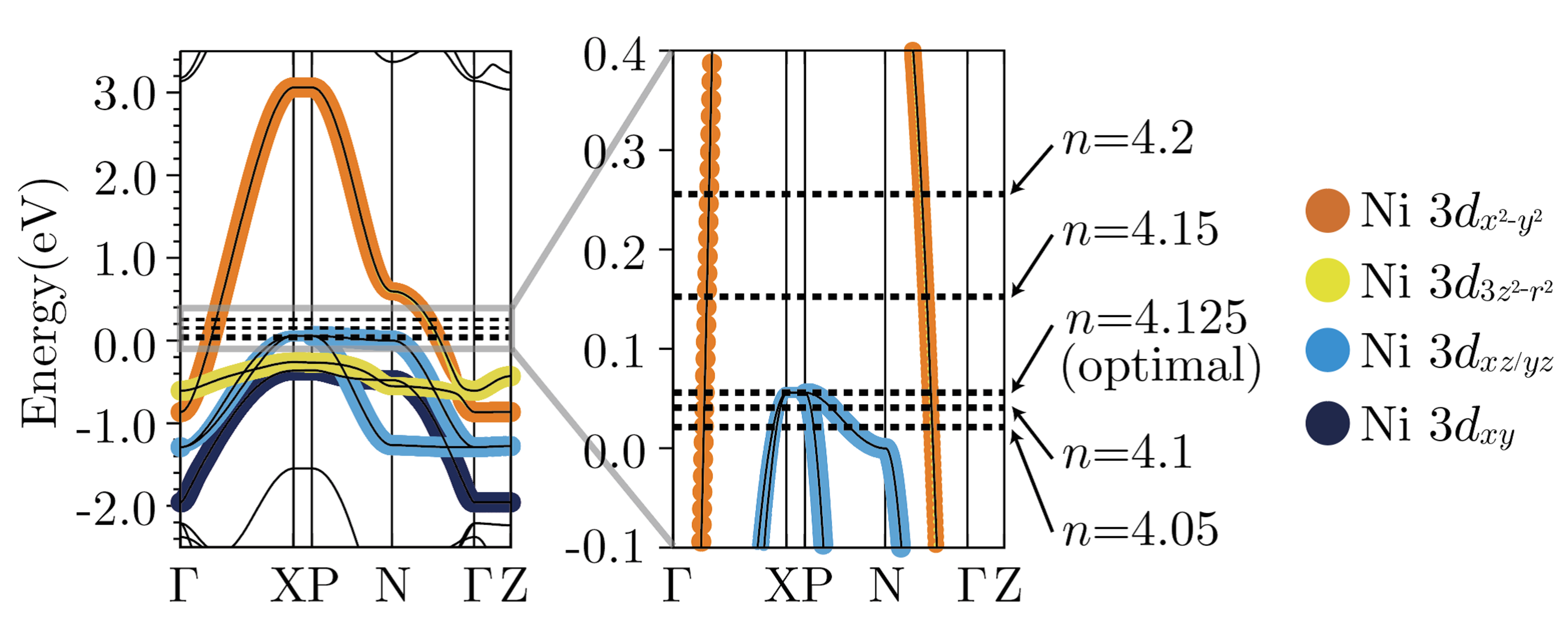}
	\caption{The band structure of the five orbital model of Ca$_2$NiO$_2$Cl$_2$, together with the position of $E_F$ for $n=4.05$, 4.1, 4.125, 4.15, and 4.2. Note that this is the bare band structure, but the Fermi surfaces themselves are not affected by the interaction since the real part of the self-energy at $\omega=0$ is subtracted in the FLEX calculation.}
	\label{figS12}
\end{figure}

\begin{figure}[h]
	\includegraphics[width=8cm]{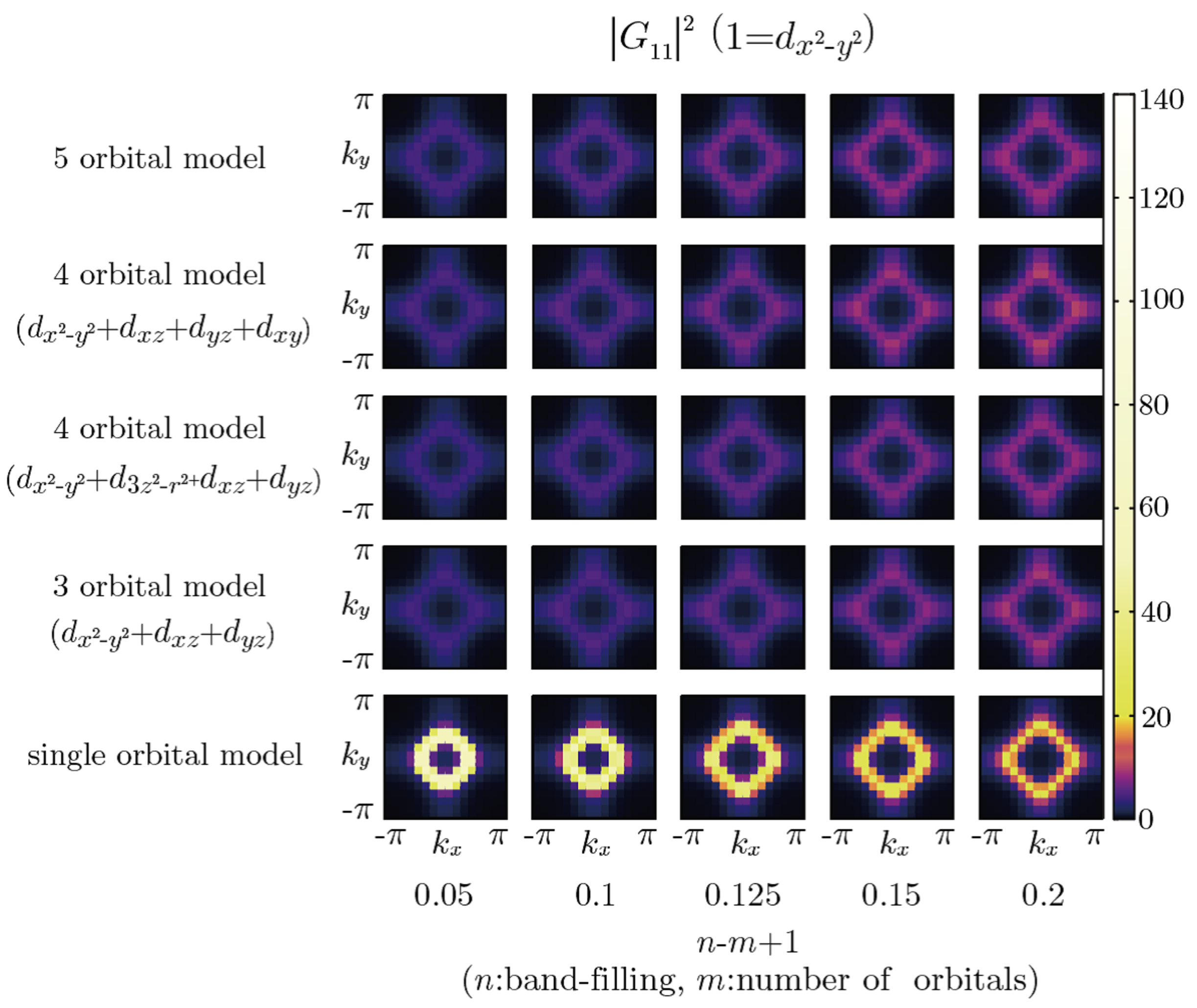}
	\caption{Contour plots of $|G_{ii}({\bm k},i\pi k_BT)|^2$ with $i=d_{x^2-y^2}$ orbital for five, four (two-types), three, and single orbital models of Ca$_2$NiO$_2$Cl$_2$ for various electron doping rates.}
	\label{figS13}
\end{figure}

In Fig.5(d) of the main text, we compared $\lambda$ of the five, four, three, and single orbital models of Ca$_2$NiO$_2$Cl$_2$ at the same electron doping rates, between 5 to 25 $\%$. Here we explain how this is justified. In Fig.\ref{figS11}, we show the contour plots of $|G_{ii}({\bm k},i\pi k_BT)|^2$, where $G_{ii}({\bm k},i\pi k_BT)$ is the orbital diagonal component of the FLEX Green's function at the lowest Matsubara frequency, for the band fillings $n=4.05$, 4.1, 4.125, 4.15, and 4.2. The ridges of these plots represent the Fermi surfaces. We also show in Fig.\ref{figS12} the position of $E_F$ in the band structure. From this, it can be seen that beyond the band filling of $n=4.125$ (12.5$\%$ doping), where $\lambda$ is maximized, the bands other than $d_{x^2-y^2}$ sink below $E_F$. In Fig.\ref{figS13}, we show the contour plot of the squared Green's function of the $d_{x^2-y^2}$ orbital for five, four, three and single orbital models, for electron doping rates between 5 and 20 $\%$.
For 15$\%$ doping or more, where the bands other than $d_{x^2-y^2}$ sink below $E_F$, the $d_{x^2-y^2}$ Fermi surface is the same for all the models as expected. Even for smaller band fillings, the $d_{x^2-y^2}$ Fermi surfaces of the multiorbital models are the same because all of these models contain the $d_{xz/yz}$ orbital, which are the only orbitals other than $d_{x^2-y^2}$ that form the Fermi surfaces. Moreover, the $d_{x^2-y^2}$ Fermi surface of the multiorbital model is also similar to that of the single orbital model because the $d_{xz/yz}$ Fermi surface is very small within this electron doping range. In a nutshell, the comparison of the multiorbital and single orbital models at the same electron doping rates is justified because the $d_{x^2-y^2}$ Fermi surfaces of these models are (nearly) the same.

\begin{figure}[h]
	\includegraphics[width=8cm]{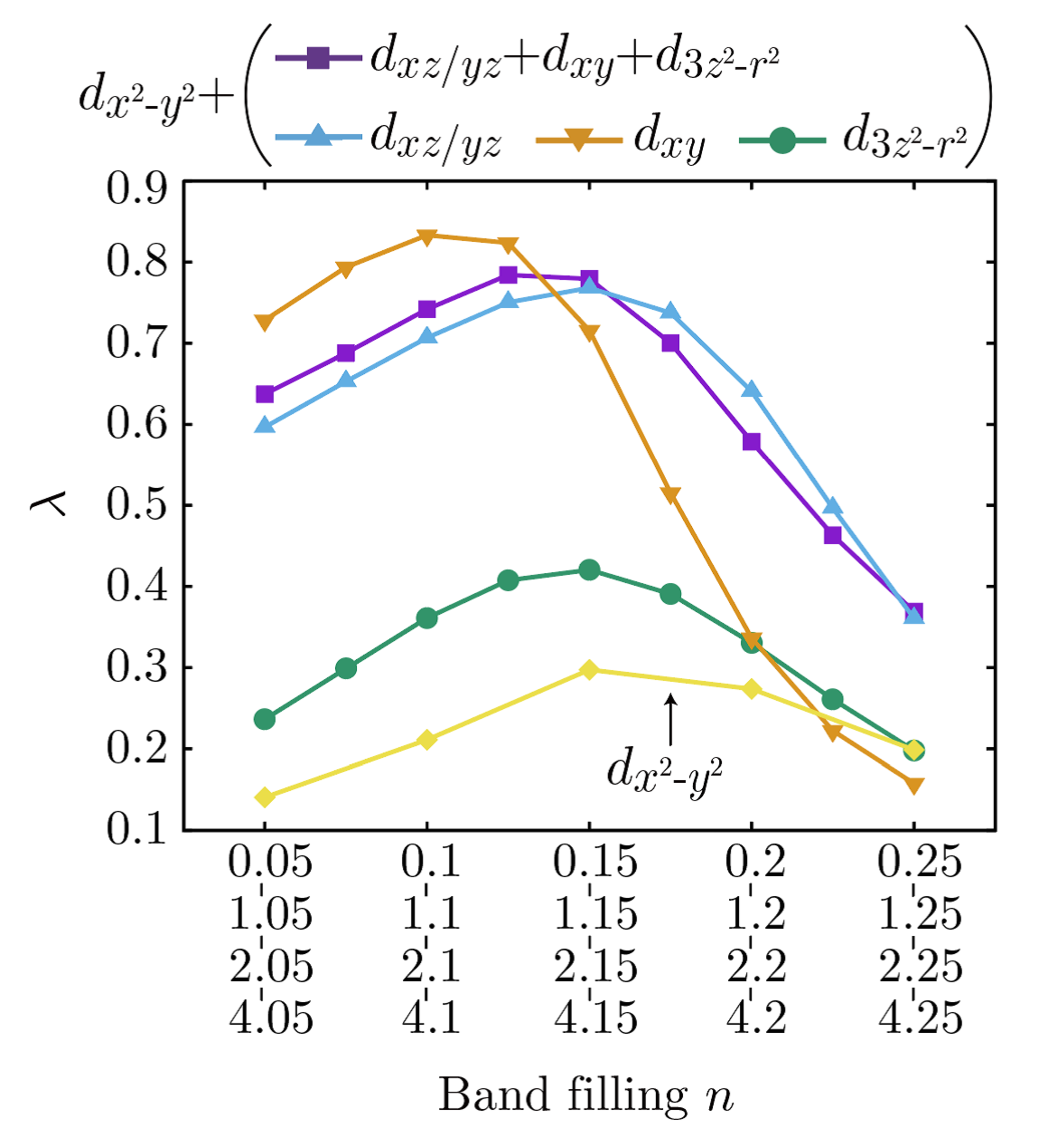}
	\caption{$\lambda$ against the band filling for various models of Ca$_2$NiO$_2$Cl$_2$.}
	\label{figS14}
\end{figure}

From the above analysis, it can also be understood why we have considered, in Fig.5(d) of the main text, only models that contain the $d_{xz/yz}$ orbitals (except for the single orbital model). Since the top of the $d_{xz/yz}$ bands has the highest energy among the bands other than $d_{x^2-y^2}$ and the Fermi level for the stoichiometric band filling of $n=4$ intersects only the $d_{x^2-y^2}$ and $d_{xz/yz}$ bands, the Fermi level of the full five orbital model with the band filling of $n(\geq 4)$ is the same as that of an $n_{\rm orb}$ orbital model with the band filling of $n-(5-n_{\rm orb})$, as far as the models that contain the $d_{xz/yz}$ orbitals are concerned. Conversely, the Fermi level of, say, a two-orbital $d_{x^2-y^2}+d_{xy}$ model with the band filling of $n-3$ is not the same as that of the full five orbital model with the band filling of $n$. Hence, such models were not considered in the main text. Nonetheless, it is interesting to study how each band would affect superconductivity if other bands were absent. In Fig.\ref{figS14}, we plot the eigenvalue of the Eliashberg equation against the band filing for two-orbital $d_{x^2-y^2}+d_{xy}$ and $d_{x^2-y^2}+d_{3z^2-r^2}$  models, in addition to the $d_{x^2-y^2}+d_{xz/yz}$ model, the single orbital $d_{x^2-y^2}$ model, and the full five orbital model. It can be seen that all of the bands can act as an incipient band that enhances $s\pm$ superconductivity compared to the single orbital case. A tendency here is that the orbitals having larger $|\Delta E|$ (level offset with respect to the $d_{x^2-y^2}$ orbital, corresponding to $2t_\perp$ in the bilayer Hubbard model) give stronger enhancement of superconductivity, and superconductivity is optimized at lower band fillings. This tendency is indeed the same as that seen in the bilayer Hubbard model\cite{Maier,Matsumoto2,DKato}.

\begin{figure}[h]
	\includegraphics[width=8.5cm]{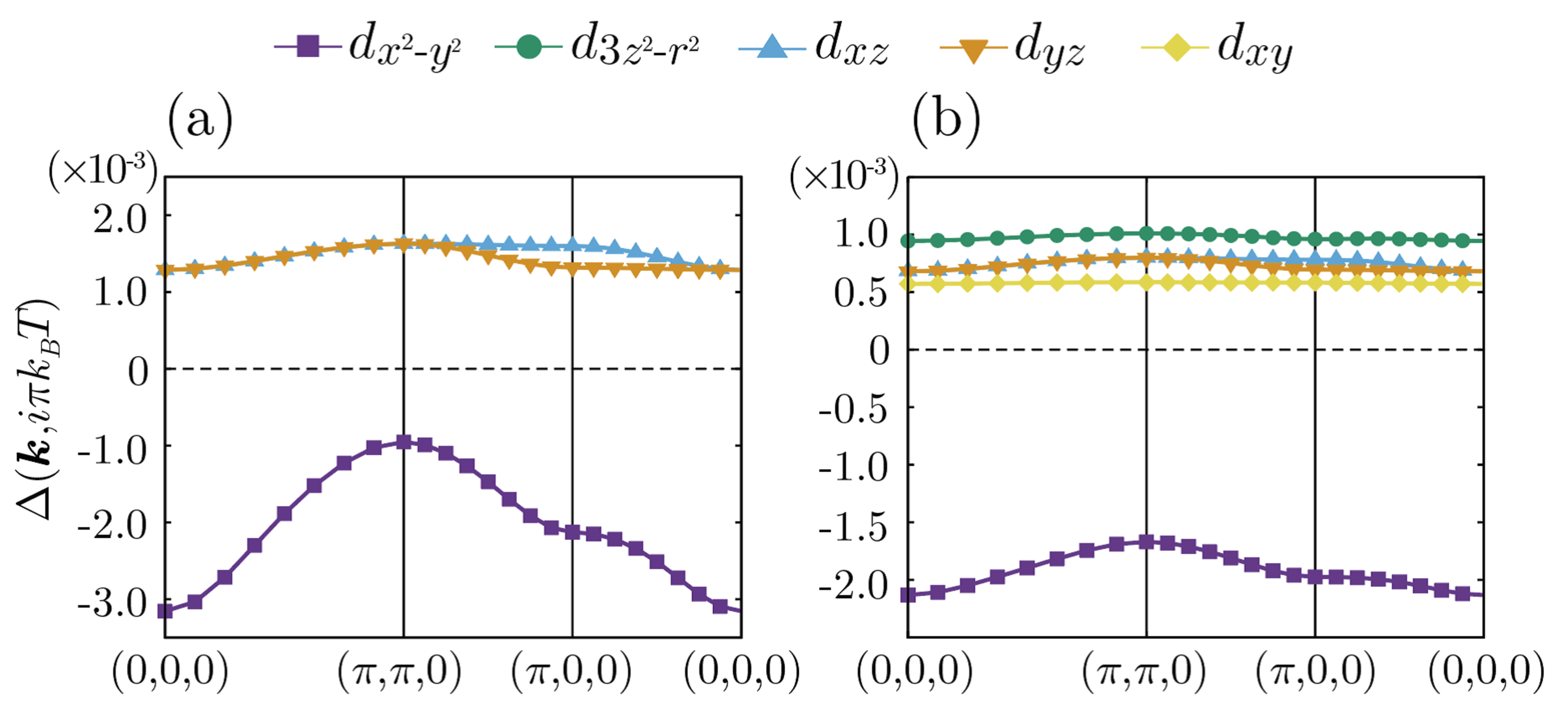}
	\caption{The gap function of (a) the $d_{x^2-y^2}+d_{xz/yz}$ model ($n=2.15$) and (b) the five orbital model ($n=4.125$) of Ca$_2$NiO$_2$Cl$_2$.}
	\label{figS15}
\end{figure}

Here we also comment on the comparison of the gap function among the multiorbital models. In Fig.\ref{figS15}(a), we show the gap funcion of the $d_{x^2-y2}+d_{xz/yz}$ model for $n=2.15$. This gap function exhibits wave number dependence for the $d_{x^2-y^2}$ orbital, which differs from the case of the bilayer Hubbard model, where the gap function barely varies against the wave number. Similar difference between the two-orbital model and the bilayer model was pointed out in Ref.\cite{Yamazaki}, and was attributed to the inequality among the interactions, i.e., $U\neq U'\neq J$.
Interestingly, as shown in Fig.\ref{figS15}(b), in the five orbital model, this wave number dependence is suppressed, and the gap function looks closer to that of the bilayer Hubbard model. The reason for this is not clear at present, and remains as an interesting future problem.

\begin{figure}[h]
	\includegraphics[width=8cm]{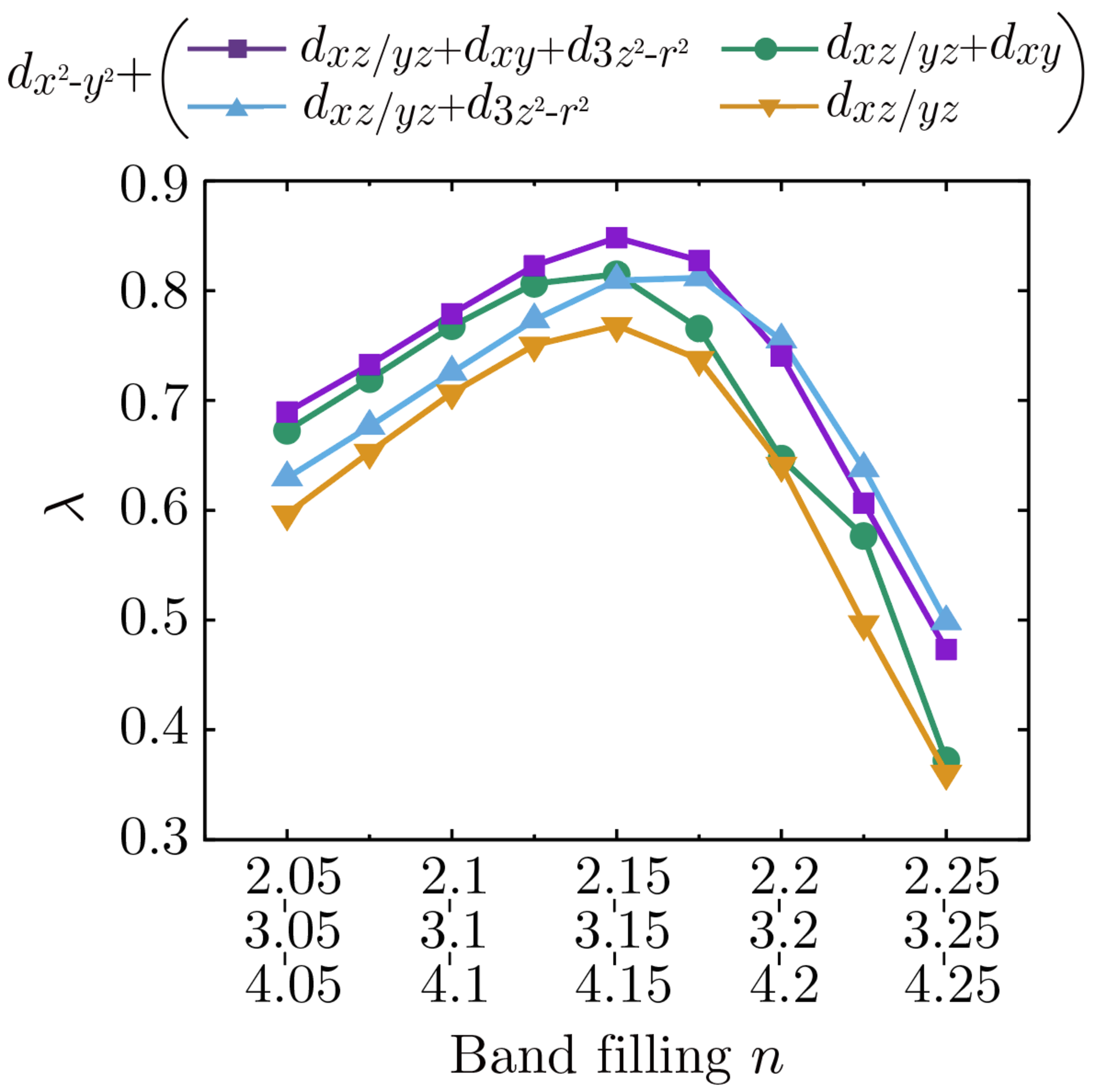}
	\caption{$\lambda$ against the band filling for various models of Ca$_2$NiO$_2$Cl$_2$. Here, the hybridization between the $d_{3z^2-r^2}$ and $d_{x^2-y^2}$ orbitals is turned off.}
	\label{figS16}
\end{figure}

\section{The effect of the orbital hybridization}
In Fig.\ref{figS14}, the enhancement of superconductivity in the $d_{x^2-y^2}+d_{3z^2-r^2}$ model appears to be rather small even considering the small $|\Delta E|$ of the $d_{3z^2-r^2}$ orbital. Moreover, as seen in Fig.5(d) of the main text, adding $d_{3z^2-r^2}$ to $d_{x^2-y^2}+d_{xz/yz}$ suppresses $\lambda$, while adding the $d_{xy}$ orbital to the $d_{x^2-y^2}+d_{xz/yz}$ model enhances it. The difference between $d_{3z^2-r^2}$ and $d_{xy}$ lies in that the former orbital hybridizes with $d_{x^2-y^2}$. In fact, as shown in Fig.\ref{figS16}, if we hypothetically turn off the hybridization, we find that adding $d_{3z^2-r^2}$ to $d_{x^2-y^2}+d_{xz/yz}$ or to $d_{x^2-y^2}+d_{xz/yz}+d_{xy}$ actually enhances $\lambda$. Similarly, it was also found in the two-orbital model in Ref.\cite{Yamazaki} that turning off the hybridization enhances superconductivity when one of the bands is incipient. As described in the first section of this supplemental material, the presence of the orbital hybridization in the two-orbital model results in an inequivalence of the two layers when transformed into the bilayer model. We believe that this should be related to the suppression of superconductivity in the presence of the hybridization, although more study on this issue is necessary for further clarification.